\documentclass[%
 reprint,
 amsmath,amssymb,
 aps,nofootinbib
]{revtex4-1}
\usepackage{bigints}
\usepackage{subfig}
\usepackage{bm}
\usepackage{placeins}
\usepackage[bottom]{footmisc}
\usepackage{graphicx}
\graphicspath{ {./figures/} }
\usepackage{dcolumn}
\usepackage{bm}
\usepackage[font={small,it}]{caption}
\usepackage[colorlinks,allcolors=blue]{hyperref}
\usepackage{lipsum}

\newcommand*{\rom}[1]{\expandafter\@slowromancap\romannumeral #1@}

\begin{document}


\title{Shape dynamics of a red blood cell in Poiseuille flow }

\author{Dhwanit Agarwal and George Biros}
\email{(dhwanit, biros)@oden.utexas.edu}
 \affiliation{Oden Institute of Computational Engineering and Sciences, University of Texas at Austin, TX 78712, USA}

\date{\today}

\begin{abstract}
{We use numerical simulations to study the dynamics of red blood cells (RBCs) in unconfined and confined Poiseuille flow. Previous numerical studies with 3D vesicles have indicated that the slipper shape observed in experiments at high capillary number can be attributed to the bistability due to the interplay of wall push and outward migration tendency at higher viscosity contrasts. In this paper, we study this outward migration and bistability using numerical simulations for 3D capsules and provide phase diagrams of RBC dynamics with and without viscosity contrast. We observe a bistability of slipper and croissants in confined Poiseuille flow with viscosity contrast as observed in experiments. }

\end{abstract}

\pacs{Valid PACS appear here}
\maketitle


\section{\label{sec:level1}Introduction}

The study of red blood cells (RBCs), a major component of mammalian blood, has been a fascinating topic of research for several decades due to its importance in quantitative understanding of microscale blood dynamics. These cells are highly deformable \cite{1, 2, 3} and lead to rich dynamics when subjected to viscous forcing. Understanding shape dynamics of a single red blood cell is a complex fluid-membrane interaction problem of fundamental importance in expanding our understanding of red blood cell suspensions \cite{4, 5, 6}. Accurate knowledge of RBC shape dynamics has also led to applications in microfluidics, for example, cell sorting in deterministic lateral displacement devices  \cite{7}.

\begin{figure}%
    \centering
    \subfloat[ ]{{\includegraphics[width=2.8cm, height=1.8cm ]{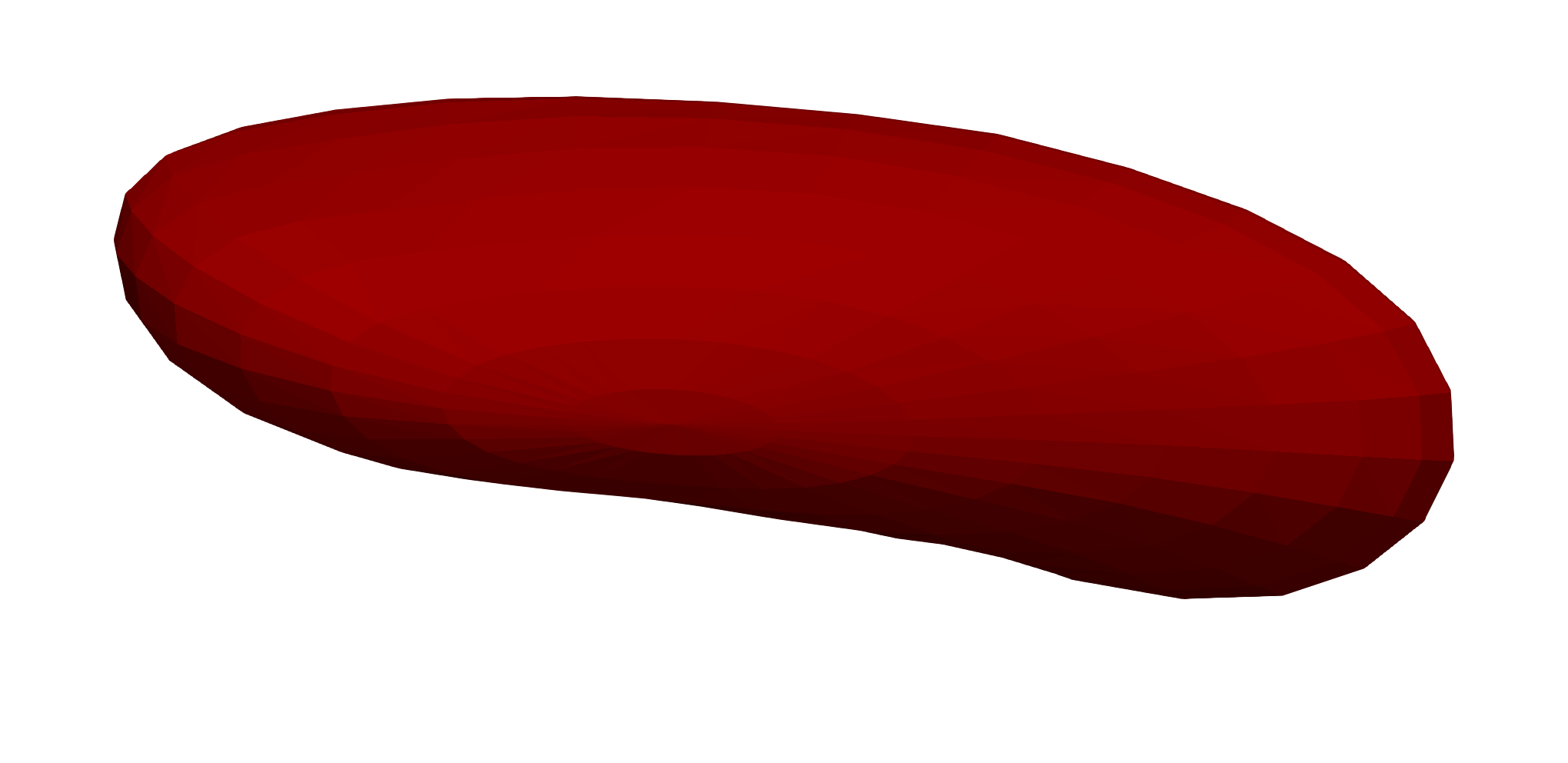} }}
    \subfloat[ ]{{\includegraphics[width=2.6cm, height=2.3cm]{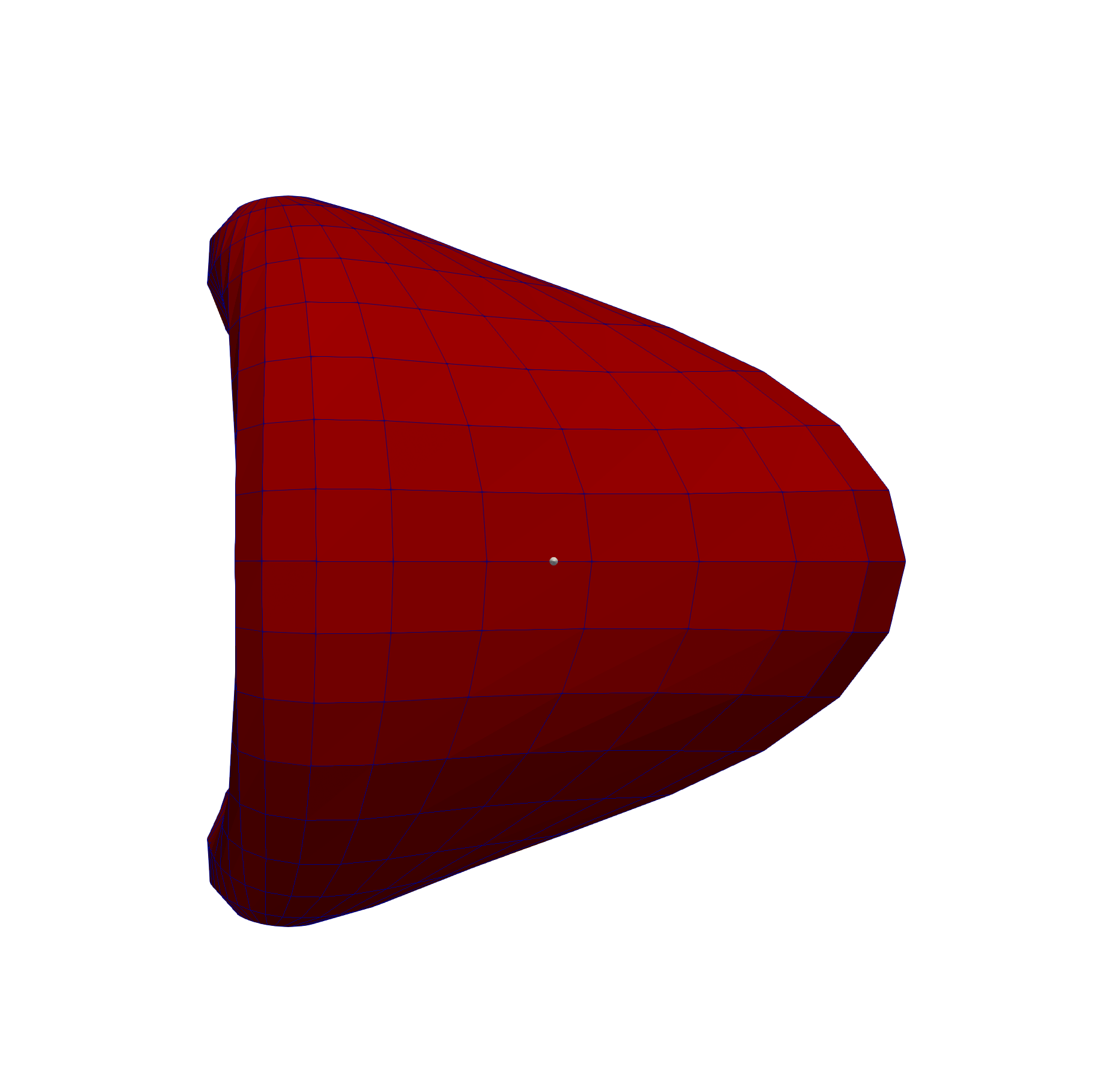} }}
    \caption{(a) Typical (a) slipper and (b) croissant observed in our simulations. }%
    \label{fig:demoshapes}%
\end{figure}

The shape dynamics of a single RBC in Poiseuille flow has been investigated both experimentally and numerically over the past few decades because of its importance in microcirculation. The dynamics depend on the elastic properties of the cell membrane, the reduced volume (denoted by $\nu$) of the RBC which is defined as the volume of the RBC over the volume of a sphere of equal surface area, the viscosity contrast (denoted by $\lambda$) between the fluid inside and outside the cell, the confinement (unconfined vs confined in a channel, and the confinement ratio, $C_{n}$, defined as the ratio of diameter of the cell over the diameter of the channel) and the properties of the imposed flow, for example, the maximum magnitude of the velocity in Poiseuille flow. The most common shapes observed in the Poiseuille flow are \textit{``slippers"} and \textit{``croissants"} (see Fig. \ref{fig:demoshapes}). The slippers are asymmetric shapes and the croissants are axisymmetric. \textit{Gaehtgens et al.} \cite{27} observed both these shapes in an experimental study.  Experiments in circular capillaries of varying diameters by \textit{ Abkarian et. al.} \cite{28} found a transition from croissants to slipper shapes as the velocity of the imposed flow is increased. However, the experiments in \cite{28} were done using a viscosity contrast $\lambda=0.3$ outside the physiological range ($\lambda=5$). \textit{Tomaiuolo et. al.} in \cite{30} reported that for confinement ratio $C_{n} \approx 0.5$ and viscosity contrast $\lambda \approx 5$, both asymmetric and axisymmetric (croissant) shapes are observed at low velocities ($ \approx 1.1$  $mm/s$). They also observed off centered slippers along with croissants in equal proportions at high velocities ( $\approx 36$ $mm/s$ ). 

Several numerical studies have modeled RBCs as a membrane enclosing a viscous Newtonian fluid and suspended in a Newtonian fluid (ignoring the cytoskeleton and viscoelastic effects). The membrane is typically endowed with bending elasticity and shear elasticity. This is referred to as the capsule model for the RBCs. Vesicles (which do not have shear resistance and are locally inextensible) have also been used in numerical studies to study the dynamics of RBCs. In 2D, there is no difference between capsules and vesicles as there is no notion of shear resistance but in 3D they have different dynamics \cite{31, 32}. Numerical studies with vesicles \cite{35, 36, 37} in Poiseuille flow have shown that a transition from asymmetric slippers to axisymmetric parachutes takes place as the flow velocity is increased. Simulations by \textit{Noguchi et. al.} in \cite{40} for a 3D capsule in confined Poiseuille flow with viscosity contrast $\lambda = 1$ and confinement around $C_{n} \approx 0.5$ found a transition from a discocyte shape to the parachute shape as the flow velocity was increased.  \textit{Fedosov et. al.} in \cite{39} used capsule model for 3D simulations of confined Poiseuille flow with viscosity contrast $\lambda = 1$ and presented a phase diagram in the parameter space of confinement ratio and capillary number. Their simulations also showed a transition from slipper to croissants as the flow velocity is increased. All these numerical results are in contrast to the experiments \cite{28, 30} where asymmetric shapes (including slippers) are observed at high flow velocity. Recent numerical studies with 3D vesicles have speculated that the existence of these slipper shapes in experiments at high flow velocities could be due to the bistability (both slipper and croissants as stable states) created due to the confinement and the outward migration of the vesicles in the presence of viscosity contrast ($\lambda > 1$) \cite{22, 37, 38}. But vesicles have no shear resistance and are quite different from RBCs. \textit{Guckenberger et. al.} in \cite{41} observed this bistability with a 3D capsule  in a confined microchannel with $\lambda=5$ and $C_{n} \approx 0.54$. However, their simulations assume the stress-free reference shape of the capsule to be a biconcave shape of reduced volume $\nu=0.65$. The stress-free reference (or ``resting") RBC shape determines the residual stress field and is an important factor in determining the shape dynamics of an RBC. Simulations of an RBC in shear flow \cite{42, 18, 26} have revealed that choosing the stress-free shape of the RBC to be an oblate near-sphere leads  to results consistent with the experiments \cite{20, 23, 26}. This is again in contrast to the previous numerical studies \cite{40, 39, 41} of 3D capsule in Poiseuille flow which assume the classical initial biconcave shape to be stress-free. 

\textit{Our Contributions:} In this study, we used a 3D biconcave capsule with a residual stress field to numerically simulate the dynamics of a single RBC in both confined and unconfined Poiseuille flow. We confirm that taking the stress-free configuration of the RBC to be an oblate near-sphere leads to results consistent with experiments. This is the first study that provides phase diagrams for a 3D capsule in unconfined and confined Poiseuille flow with viscosity contrast and initial conditions (initial shape, residual stress field) relevant to the experiments. We also provide a qualitative explanation of the bistability observed in confined flow using the results of outward migration from our unconfined flow phase diagrams. Our results extend the observations made in \cite{22} to 3D capsules and offer a possible explanation of the observation of slipper shapes at high capillary numbers in experiments\cite{29, 41}. 

\section{\label{sec:method}Problem Formulation and Methodology}
Our formulation follows the three-dimensional vesicle formulation given in \cite{21, 22} but now we include shear resistance. We assume a neo-Hookean constitutive law for shear. We assume that the RBC is perfectly locally inextensible so the tension is a Lagrange multiplier that enforces this constraint. 
Table \ref{symbols} summarizes the notation used in the paper.

\begin{table}[h!]
 \begin{tabular}{|c c|}
 \hline
 Symbol & Definition\\ [0.5ex] 
 \hline\hline
  $\gamma$ & Boundary of capsule \\
 \hline
 $\Gamma$ & Fixed rigid boundary  \\
 \hline
 $\mathbf{S}_{\gamma}$ &  The single-layer Stokes operator 
over  $\gamma$ \\
 \hline
 $\mathbf{D}_{\gamma}$ &  The double-layer Stokes operator
over  $\gamma$ \\
 \hline
 $\boldsymbol{u}$ & velocity \\
 \hline
 $\boldsymbol{u}_{\infty}$ & Background velocity \\
 \hline
 $\lambda$ & Viscosity contrast = $\mu_\mathrm{i} / \mu_\mathrm{e}$\\
 \hline
 $P $ & Pressure\\
 \hline
 $\boldsymbol{n}$ & Normal to capsule surface \\
 \hline
 $\mu_\mathrm{e}$ & Viscosity of ambient fluid\\
 \hline
 $\lambda$ & Viscosity contrast = $\mu_\mathrm{i} / \mu_\mathrm{e}$\\
 \hline
 $t $ & Time\\
 \hline
 $\sigma$ & Tension\\
 \hline
 $\tau$ & In-plane shear stress tensor\\
 \hline
 $\mathbf{f}$ & Interfacial force\\
 \hline
 $\mathbf{f_{b}}$ & Bending force\\
 \hline
 $\mathbf{f}_{\sigma}$ & Tensile force\\
 \hline
 $\mathbf{f}_{s}$ & Shear force\\
 \hline
 
 $H$ & Mean curvature of capsule \\
 \hline
 $K$ & Gaussian curvature of capsule \\
 \hline
 $\kappa_{b}$ & Bending modulus of capsule membrane \\
 \hline
 $E_{s}$ & Elastic shear modulus of capsule membrane \\
 \hline
 $E_{D}$ & Elastic dilatation modulus of capsule membrane \\
 \hline
 $\bm{\eta}$ & Double layer density on $\Gamma$\\
 \hline
 $\omega$ & Volume enclosed by $\gamma$\\
 \hline
 $\Omega$ & Volume of interest\\
 \hline
 $\mathbf{F}$ & Deformation gradient\\
\hline
 $\mathbf{F}_{s}$ & Surface deformation gradient\\
 \hline
 $I_{1}, I_{2}$ & Strain invariants for capsule surface\\
 \hline
 $\Delta t$ & Time step\\
 \hline
 $ t^{n}$ & Time after $n$ time steps, \emph{i.e.}, $n\Delta t$\\
 \hline
 $R_{0}$ & Radius of capsule\\
 \hline
 $D$ & Diameter of confining channel\\
 \hline
 $V$ &  Volume of capsule\\
 \hline
 $A$ & Area of capsule\\
 \hline
 $\alpha_{0}$ & Curvature of Poiseuille flow\\
 \hline
 $\dot{\gamma}$ & Shear rate of shear flow\\
 \hline
 $\nu$ & Reduced volume of capsule\\
 \hline
  $C_{a}$ & Bending-force capillary number\\
 \hline
  $C_{k}$ & Shear-force capillary number\\
 \hline
 $\chi$ & Bending stiffness ratio\\
 \hline
 $C_{n}$ & Confinement ratio\\
 \hline
 $\psi$ &  Capsule's terminal inclination angle\\
 \hline
 $Y_{0}$ & Initial lateral position of capsule from centerline\\
 \hline
 \end{tabular}
 \captionof{table}{Index of different symbols}\label{symbols}
\end{table}

%

\begin{figure}%
    \centering
    \subfloat[Unconfined flow \label{fig:schematic_uncon}]{{\includegraphics[width=3.2cm, height=2.3cm]{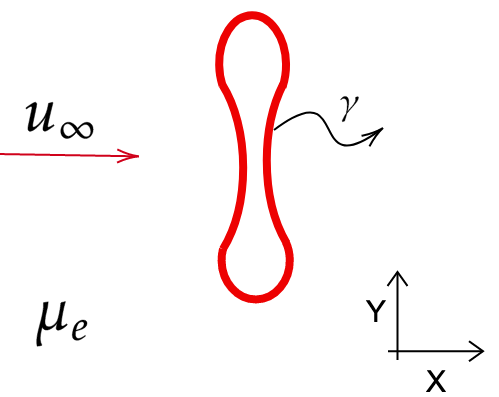} } }%
    \qquad
    \subfloat[Confined flow \label{fig:schematic_con}]{{\includegraphics[width=4.4cm, height=2.7cm]{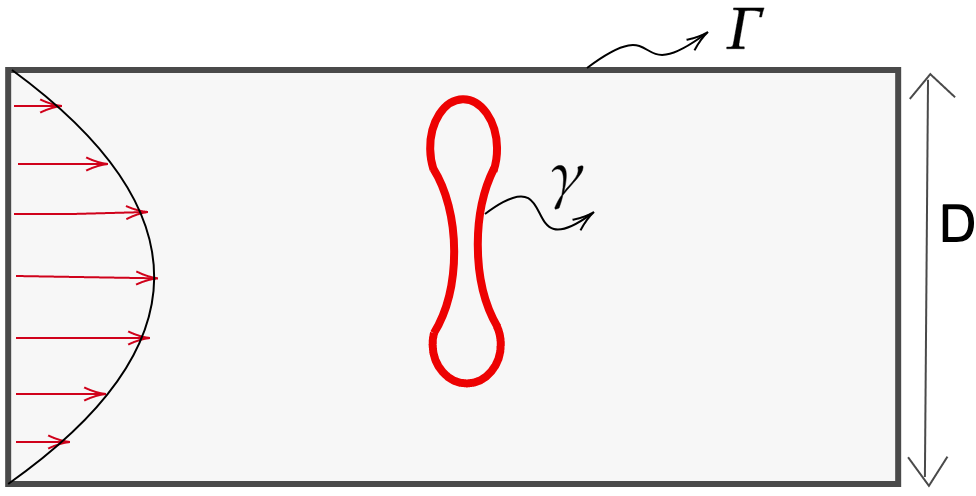} }}%
    \caption{(a) Problem setup for unconfined flow. Biconcave red boundary $\gamma$ is the capsule membrane filled with fluid of viscosity $\mu_{i}$.  Fluid outside has a viscosity of $\mu_{e}$. $u_{\infty}$ is the imposed background velocity. (b) Problem setup for confined flow (side view).  Red boundary is the capsule membrane $\gamma$ filled with fluid of viscosity $\mu_{i}$. $\Gamma$ is the fixed rigid enclosing boundary, which models a channel of circular cross-section with axis  parallel to x-axis. $D$ is the diameter of the circular cross section. We impose the parabolic flow imposed as shown by red arrows. Grey region is filled with fluid of viscosity $\mu_{e}$. }%
    \label{fig:schematic}%
\end{figure}

\subsection{\label{sec:unconfinedmodel} Unconfined flow formulation}

The formulation for a capsule in unconfined flow is given below. A representation of the setup is shown in Fig. \ref{fig:schematic_uncon}. The Reynolds number is low so the inertial effects can be ignored. The PDE formulation of the flow is as follows:

\begin{align} 
-\mu(\boldsymbol{x}) \Delta \boldsymbol{u}(\boldsymbol{x}) + \nabla P(\boldsymbol{x}) = 0   \textbf{    } \forall \mathbf{x} \in \mathbb{R}^{3}\backslash \gamma  \label{eq1},\\
div (\boldsymbol{u}(\boldsymbol{x}) )= 0  \textbf{     } \forall \mathbf{x} \in \mathbb{R}^{3}\backslash \gamma  \label{eq2},\\
[[ -P\mathbf{n} + (\nabla \boldsymbol{u} + \nabla \boldsymbol{u}^{T})\mathbf{n}]] = \mathbf{f}  \textit{ on  }  \gamma \label{eqjump},\\
\frac{\partial \boldsymbol{x}}{\partial t} = \boldsymbol{u}(\mathbf{x})  \textbf{     }  \forall  \boldsymbol{x} \in \gamma \label{eq3}, \\
 \boldsymbol{u}(\boldsymbol{x})  \rightarrow \boldsymbol{u}_{\infty} (\boldsymbol{x}) \textit{ as   }  {||\boldsymbol{x}||} \rightarrow \infty \label{eqinf}, \\
 div_{\gamma} (\boldsymbol{u}(\boldsymbol{x}) )= 0  \textbf{     } \forall \boldsymbol{x} \in  \gamma  \label{eqsurfdiv}, \\
 \mathbf{f} = \mathbf{f}_{b} + \mathbf{f_{\sigma}} + \mathbf{f_{s}}, \label{eqforce} \\
\mathbf{f_{b}}(\boldsymbol{x}) = -4 \kappa _{b} [\Delta_{\gamma}H + 2H(H^{2} - K))] \mathbf{n}, \textbf{     } \forall \boldsymbol{x} \in  \gamma, \label{eqbending} \\
 \mathbf{f_{\sigma}}(\boldsymbol{x}) = \sigma \Delta_{\gamma}\boldsymbol{x} + \nabla_{\gamma} \sigma \textit{} \textbf{     } \forall \boldsymbol{x} \in  \gamma, \label{eqtension} \\
 \mathbf{f_s}(\boldsymbol{x}) = div_{\gamma} \mathbf{\tau} \textbf{     } \forall \boldsymbol{x} \in  \gamma, \label{eqshear}
\end{align}
where $\gamma$ is the capsule membrane, $\mathbf{u}(\boldsymbol{x})$ is the velocity of  the fluid and  $P(\boldsymbol{x})$ is the pressure. The viscosity $\mu$ is given by the piecewise function 
\[ \mu(\boldsymbol{x}) = \begin{cases} 
      \mu_\mathrm{i} & \text{ if } \boldsymbol{x} \in \omega, \\
       \mu_\mathrm{e} & \text{ if } \boldsymbol{x} \in \mathbb{R}^{3} \backslash \omega. \\
     \end{cases}
\]
$[[q]]$ denotes the jump of the quantity $q$ across the capsule membrane and $\mathbf{n}$ is the outward unit normal to the membrane. Equation (\ref{eqjump}) is the balance of momentum on membrane, which requires the surface traction jump to be equal to the total force (denoted by $\mathbf{f}$) exerted by the interface onto fluid. Equation (\ref{eq3}) enforces no-slip boundary condition on capsule membrane and equation (\ref{eqinf}) sets the far field velocity to be the background velocity. The viscosity contrast, $\lambda$, is defined to be $\lambda:=\frac{\mu_{i}}{\mu_{e}}$.  Equation (\ref{eqsurfdiv}) enforces the inextensibility of the capsule membrane which is mathematically equivalent to requiring that the surface divergence of velocity should vanish on the capsule membrane. 
 Equation (\ref{eqforce}) gives the elastic force $\mathbf{f}$ due to the capsule membrane elasticity which comprises of a bending component $\mathbf{f}_{b}$, a tension component $\mathbf{f}_{\sigma}$ and a shear component $\mathbf{f}_{s}$. 
The expressions for the bending and the tension components are given in Equations (\ref{eqbending}) and (\ref{eqtension}) respectively (please refer to \cite{10, 21, 22} for details), 
 where $\kappa_{b}$ is membrane's bending modulus, $H$ and $K$  are the mean and Gaussian curvature respectively of the membrane at $\mathbf{x} \in \gamma$ and $\sigma$ is the tension at the membrane point $\boldsymbol{x}$. Equation (\ref{eqshear}) gives the shear force, denoted by $\mathbf{f}_{s}$. It is equal to the surface divergence of the symmetric part of the in-plane shear stress tensor $\mathbf{\tau}$ \cite{8}. We ignore the anti-symmetric part of $\mathbf{\tau}$, which arises due to bending moment, since we have already included the bending force albeit in a different form. The evaluation of $\mathbf{\tau}$ is discussed later in this section. 
With this shear force at our disposal, following \cite{8, 9} the capsule dynamics equations can be written in integral form for $\boldsymbol{x} \in \gamma$
\begin{align}
\alpha \boldsymbol{u}(\boldsymbol{x}) =  \boldsymbol{u}_{\infty} (\boldsymbol{x})  + \mathbf{S}_{\gamma}[\mathbf{f_{b}} + \mathbf{f_{\sigma}} + \mathbf{f_{s}}] (\boldsymbol{x}) + \mathbf{D}_{\gamma}[\boldsymbol{u}] (\boldsymbol{x}) \label{syseq1},\\
div_{\gamma}(\boldsymbol{u} (\boldsymbol{x})) = 0 \label{syseq2},\\
\frac{\partial \boldsymbol{x}}{\partial t} = \boldsymbol{u} ({\boldsymbol{x}}) \label{syseq3},
\end{align}
where $\alpha := (1 + \lambda)/ 2$. The single layer convolution integral is defined as  $\mathbf{S}_{\gamma}[\mathbf{f}] (\boldsymbol{x}) := \bigints_{\gamma} S_{0}(\boldsymbol{x}, \mathbf{y}) \mathbf{f(\mathbf{y})} d\gamma $, with
\begin{align*}
S_{0}(\boldsymbol{x}, \mathbf{y} ) = \frac{1}{8 \pi \mu} \frac{1}{||\mathbf{r}||} (I + \frac{\mathbf{r} \otimes \mathbf{r}}{||\mathbf{r}||^{2}}),
\end{align*}
where $\mathbf{r}:= \boldsymbol{x} - \mathbf{y}$, $I$ is the identity operator, $\otimes$ is tensor product and $||\cdot||$ is the Euclidean norm.  The double layer convolution integral is defined as $\mathbf{D}_{\gamma}[\mathbf{f}] (\boldsymbol{x}) := (1-\lambda)\bigints_{\gamma} D_{0}(\boldsymbol{x}, \mathbf{y}) \mathbf{f(\mathbf{y})} d\gamma $, with 
  \begin{align*}
  D_{0}(\boldsymbol{x}, \mathbf{y} ) = \frac{-3}{4 \pi}(\mathbf{r} \cdot \mathbf{n})\frac{\mathbf{r} \otimes \mathbf{r}}{||\mathbf{r}||^{5}}.
  \end{align*}

\textit{In-plane shear stress tensor $\mathbf{\tau}$:} The shear stress tensor, $\tau$, is a function of the relative surface deformation gradient. Now we discuss in detail the formulation of the shear force in this work. Let $\gamma$ denote the capsule membrane in the current configuration and $\gamma_{r}$ denote the membrane in the reference configuration. For $\mathbf{x_{r}} \in \gamma_{r}$, let $\phi$ denote the deformation map from the reference configuration to the current configuration such that $\phi(\mathbf{x_{r}}) = \mathbf{x}$, where $\mathbf{x} \in \gamma$. The Cartesian deformation gradient $\mathbf{F(\mathbf{x_{r}})}$  for 3D continuum is defined as follows:
\begin{align}
\mathbf{F(x_{r})} = \frac{\partial \mathbf{x}}{\partial \mathbf{x_{r}}}. 
\end{align}
Let $\mathbf{n_{r}}$ denote the normal to the membrane surface in the reference state. Then the surface deformation gradient is the surface projection of the deformation gradient, given by $\mathbf{F_{s}(\mathbf{x_{r}})} := \mathbf{F}(\mathbf{I} - \mathbf{n_{r}\mathbf{n^{T}}})$. Let $\mathbf{a}_{1r}$ and $\mathbf{a}_{2r}$ be two material fibers tangential to the surface in the reference state $\gamma_{r}$ at $\mathbf{x_{r}}$. If  these tangential fibers are transformed to tangential fibers $\mathbf{a_{1}} $ and $\mathbf{a_{2}}$ in the current state at position $\mathbf{x} \in \gamma$ with $\mathbf{x} = \phi(\mathbf{x_{r}})$, then we write,
\begin{align}
\mathbf{F_{s}} \mathbf{a_{1r}} = \mathbf{a_{1}} \textit{ and }\mathbf{F_{s}} \mathbf{a_{2r}} = \mathbf{a_{2}} \label{eqsurfdeform}.  
\end{align}
If $\mathbf{a_{1r}}$ and $\mathbf{a_{2r}}$ are orthogonal, then $\mathbf{F_{s}}$ can be directly written as $\mathbf{F_{s}}(\mathbf{x_{r}}) = \mathbf{a}_{1} \otimes (\mathbf{a}_{1r}/ ||\mathbf{a}_{1r}||^{2}) + \mathbf{a}_{2} \otimes (\mathbf{a}_{2r}/ ||\mathbf{a}_{2r}||^{2})$. We can use the deformation map $\phi$ to write the surface deformation gradient in the current configuration as  $\tilde{\mathbf{F}}_{s}(\mathbf{x}) = \mathbf{F_{s}}(\phi(\mathbf{x_{r}}))$. 
From $\mathbf{\tilde{F}_{s}}$, we follow \cite{8} to construct the left Cauchy-Green deformation tensor 
\begin{align}
    \mathbf{V}^{2} =  \mathbf{\tilde{F}_{s}} \mathbf{\tilde{F}_{s}}^{T}.
\end{align}
Let $\lambda_{1, 2}^{2}$ be the two non-zero eigenvalues of $\mathbf{V}^{2}$ with the the third being zero. The surface strain-energy function $W$ depends on the
surface deformation gradient through the strain invariants,
\begin{align}
    I_{1} = \lambda_{1}^{2} + \lambda_{2}^{2} - 2, \textit{   } I_{2} = \lambda_{1}^{2}\lambda_{2}^{2} - 1.
\end{align}
Skalak et al.  proposed the following strain energy function for the membrane (please refer to \cite{8} for details):
\begin{align}
    W = \frac{E_{s}}{4}(0.5 I_{1}^{2} + I_{1} - I_{2}) + \frac{E_{D}}{8}I_{2}^{2},
\end{align}
where $E_{s}$ is the elastic shear modulus and $E_{D}$ is the dilatation modulus to maintain surface dilatation of unity. The in-plane Cauchy stress tensor is then given by, 
\begin{align}
    \mathbf{\tau} = \frac{E_{s}}{2J_{s}}(I_{1} + 1) \mathbf{V}^{2} + \frac{J_{s}}{2} (E_{D}I_{2} - E_{s}) \mathbf{P},
\end{align}
where $\mathbf{P} = \mathbf{I} - \mathbf{n}\mathbf{n}^{T}$ is the surface projection tensor with $\mathbf{n} =  \mathbf{n(x)}$ being the normal in the current configuration of the membrane. We define $J_{s} = \lambda_{1}\lambda_{2}$. Since we are imposing inextensibility of membrane as a separate constraint, we take $E_{D} = 0$ in our formulation.  The shear force $\mathbf{f_{s}(x)}$ is now given by $\mathbf{f_{s}(x)} = div_{\gamma}\mathbf{\tau}$.\\

\textit{Discretization:} We use spherical harmonics discretization for the capsule surface $\gamma$ and the functions defined on $\gamma$. We use the singular quadratures described in \cite{34,21} to evaluate the integrals. The system of equations (\ref{syseq1})--(\ref{syseq3}) is then solved using a linearly semi-implicit scheme \cite{22} for the velocity $\mathbf{u}$ and tension $\sigma$. Let $t^{n}$ be the time elapsed after $n$ time steps of step size $\Delta t$ such that $t^{n} = n\Delta t$. Let $\mathbf{x}^{n}$ denote the capsule position at time $t^{n}$. The discretized system of equations is given below:
\begin{align}
\alpha \boldsymbol{u}^{n+1} &=  \boldsymbol{u}_{\infty}^{n}  +  \mathbf{S}_{\gamma}[\mathbf{f_{b}(\mathbf{x}^{n}) + f_{b}^{'}(\mathbf{x}^{n})\boldsymbol{u}^{n+1}} \Delta t  \notag \\ 
&+ \mathbf{f_{s}(\mathbf{x^{n}})} + \mathbf{f_{\sigma}}(\sigma^{n+1}( \boldsymbol{x^{n}})]  + \mathbf{D}_{\gamma}[\boldsymbol{u^{n+1}}] ,    \label{dnewsyseq1}\\
div_{\gamma}(\boldsymbol{u}^{n+1}) &= 0 \label{dnewsyseq2},\\
\frac{\mathbf{{x_{0}}}^{n+1} - \mathbf{x}^{n}}{\Delta t} &= \boldsymbol{u}^{n+1} \label{dnewsyseq3}.
\end{align}
Here $\mathbf{{x_{0}}}^{n+1}$ denotes the membrane position at time $t^{n+1}$ before reparameterization. To maintain the mesh quality for long timescale simulations, we use the reparameterization scheme described in \cite{34,21} to move the points $\mathbf{{x_{0}}}^{n+1}$ along the surface. The reparameterization scheme moves the discretization points $\mathbf{x_{0}}^{n+1}$ to $\mathbf{x}^{n+1}$. Thus, the reparameterization scheme changes the material points we are tracking after each time step. To calculate the surface deformation gradient $\mathbf{\tilde{F}_{s}}^{n+1}$ at $\mathbf{x}^{n+1}$ using Eq. (\ref{eqsurfdeform}), we will require the tangential fibers at these new material points in the reference state, which are not available. Hence, we propose a local linear Taylor approximation to approximate these deformation gradients. Let the surface deformation gradient at $\mathbf{x}^{n}$ be $\mathbf{\tilde{F}_{s}}^{n}$. Let $\mathbf{a_{1}}^{n}$ and $\mathbf{a_{2}}^{n}$ be the tangential fibers at $\mathbf{x}^{n}$ that get transformed to $\mathbf{a_{1, 0}}^{n+1}$ and $\mathbf{a_{2, 0}}^{n+1}$ at $\mathbf{x_{0}}^{n+1}$. Then the surface deformation gradient $\mathbf{\tilde{F}_{s, 0}}^{n+1}$ at the discretization points $\mathbf{x_{0}}^{n+1}$ is given by, 
\begin{align}
\mathbf{\tilde{F}_{s,0}}^{n+1} = \Delta \mathbf{\tilde{F}_{s}} \mathbf{\tilde{F}_{s}}^{n}, 
\end{align}
 where $\Delta \mathbf{\tilde{F}_{s}}$ is the relative surface deformation gradient at $\mathbf{x_{0}}^{n+1}$ from time $t^{n}$ to $t^{n+1}$, given by $\Delta \mathbf{\tilde{F}_{s}} \mathbf{a_{i}}^{n} = \mathbf{a_{i,0}}^{n+1} \textit{, } i=1,2$. Then we use the local Taylor approximation to get $\mathbf{\tilde{F}_{s}}^{n+1}$ at $\mathbf{x}^{n+1}$, given by,  
\begin{align}
    \mathbf{\tilde{F}_{s}}^{n+1} \approx \mathbf{\tilde{F}_{s,0}}^{n} + \nabla_{\gamma} \mathbf{\tilde{F}_{s,0}}^{n} (\mathbf{x}^{n+1} - \mathbf{x_{0}}^{n+1}).
\end{align}
\subsection{\label{sec:confinedmodel} Confined flow formulation}
We define the capsule characteristic radius, denoted by $R_{0}$, as the radius of the sphere that has the same area as the capsule. To model the flow of a capsule in the confined Poiseuille flow, we create a channel with length larger than the capsule radius and a circular cross-section. The length of the channel is eight times the capsule radius $R_{0}$. The axis of the channel is parallel to $x$-axis (refer to Fig. \ref{fig:schematic_con}). To account for the confinement, we follow the scheme in \cite{22} and add a capsule-wall interaction term to the equation (\ref{syseq1}). We append one more equation (\ref{consyseq3}) for the calculation of the unknown double layer density $\bm{\eta}$ on the fixed rigid boundary $\Gamma$. The formulation becomes: 
 \begin{align}
\alpha \boldsymbol{u}(\mathbf{x}) &=   \mathbf{S}_{\gamma}[\mathbf{f_{b}} + \mathbf{f_{\sigma}} + \mathbf{f_{s}}] (\mathbf{x}) + \mathbf{D}_{\gamma}[\boldsymbol{u}] (\mathbf{x})  +  \mathbf{D}_{\Gamma}[\bm{\eta}](\mathbf{x}) ,  \textit{           }    \label{consyseq1}  \\
 div_{\gamma}(\boldsymbol{u} (\mathbf{X})) &= 0 \label{consyseq2} \textit{      } \forall \mathbf{x} \in  \gamma ,  \\
\mathbf{U}(\mathbf{x}) &= -\frac{1}{2}\bm{\eta}(\mathbf{x}) + \mathbf{S}_{\gamma}[\mathbf{f_{b}} + \mathbf{f_{\sigma}} + \mathbf{f_{s}}] (\mathbf{x}) + \mathbf{D}_{\gamma}[\boldsymbol{u}] (\mathbf{x}) \notag \\
&\phantom{{}=1} +  \mathbf{D}_{\Gamma}[\bm{\eta}](\mathbf{x}) + \mathbf{N}_{0}[\bm{\eta}](\mathbf{x})    \textit{            } \forall \mathbf{x} \in \Gamma,  \label{consyseq3}\\
\frac{\partial \mathbf{x}}{\partial t} &= \boldsymbol{u} ({\mathbf{x}})   \textit{            } \forall \mathbf{x} \in \gamma,   \label{consyseq4}
\end{align}  
where $\mathbf{N}_{0}[\mathbf{\bm{\eta}}](\mathbf{x})  = \mathbf{n}(\mathbf{x})\bigints_{\Gamma} \left( \mathbf{n}(\mathbf{y})\cdot \bm{\eta}(\mathbf{y})\right) ds(\mathbf{y})$ and $\mathbf{U}(\mathbf{x})$ is the given velocity of rigid enclosing boundary at $\mathbf{x} \in \Gamma$. We solve the system of equations (\ref{consyseq1}-\ref{consyseq2}) for $\mathbf{u}$ and $\sigma$ as in the unconfined case. We then use the obtained $\mathbf{u}$ and $\sigma$ in (\ref{consyseq3}) to solve for double layer density $\bm{\eta}$ on $\Gamma$. Finally, equation (\ref{consyseq4}) is discretized as  $\mathbf{x}_{new} = \mathbf{u}\Delta t + \mathbf{x}_{old}$ to solve for new vesicle position $\mathbf{x}_{new}$. To avoid the effect of finite length of the channel, after each time step we translate the vesicle so that the $x$-coordinate of the center of vesicle coincides with the $x$-coordinate of the center of the channel.

\section{\label{sec:simulation} Parameters and validation}

\subsection{\label{sec:parameters} Non-dimensional parameters}
To study the shape dynamics of capsules, the following non-dimensional parameters are of key importance:

\begin{itemize}
  \item \textit{Viscosity contrast:} The viscosity contrast, denoted by $\lambda$, is the ratio of dynamic viscosity of fluid inside the capsule to the viscosity of the suspending fluid, i.e., $\lambda := \frac{\mu_{i}}{\mu_{e}}$.  
  \item \textit{Reduced volume:} The reduced volume of the capsule, denoted by $\nu$, is defined as the ratio of the volume of capsule to the sphere with the same area as the capsule.  It is given by
 \begin{align} 
  \nu := 6 \pi^{\frac{1}{2}}V A^{\frac{-3}{2}},
  \end{align}
  where $V$ is the volume of the capsule and $A$ is the surface area of the capsule. In most of our simulations, $\nu=0.65$. In the validation results, we compare with the literature and use other values of $\nu$. 
  \item \textit{Capillary number:} The bending-force capillary number, denoted by $C_{a}$, is defined as the ratio of the shear rate over the bending energy of the membrane. For shear flow given by $\boldsymbol{u}_{\infty} = (\dot{\gamma}y, 0, 0)$, 
  \begin{align}
  C_{a} := \frac{\dot{\gamma} \mu_{e} R_{0}^{3}}{\kappa_{b}},
  \end{align}
  where $R_{0}$ is the characteristic radius of the capsule and $\dot{\gamma}$ is the shear rate. 
  The shear-force capillary number, denoted by $C_{k}$, is defined as the ratio of the shear rate over the shear energy of the membrane. For shear flow, 
  \begin{align}
  C_{k}:= \frac{\dot{\gamma} \mu_{e} R_{0}}{E_{s}}. 
  \end{align}
 For Poiseuille flow with cross section diameter $D$ given by $\boldsymbol{u}_{\infty}(x, y, z)= \alpha_{0} (\frac{D^{2}}{4} - y^{2} - z^{2})$, we define the shear-force capillary number as:
 \begin{align}
 C_{k}:=\frac{\alpha_{0} \mu_{e} R_{0}^{2}}{E_{s}}.
 \end{align}
 
  \item \textit{Bending stiffness ratio:} The dimensionless bending stiffness ratio, denoted by $\chi$, is a measure of the ratio of bending energy to shear elastic energy of the membrane. It is given by 
  \begin{align}
  \chi := \frac{\kappa_{b}}{E_{s} R_{0}^{2}}.
  \end{align}
  
  \item{\textit{Confinement ratio} (relevant only for confined flow):} The confinement ratio, denoted by $C_{n}$, is the ratio of the capsule diameter to the diameter of the enclosing rigid confinement. It is given by 
  \begin{align}
  C_{n} := \frac{2R_{0}}{D},
  \end{align} 
  where $D$ is the diameter of the rigid wall. 
\end{itemize}

\subsection{\label{sec:validation} Validation}
\begin{figure*}%
    \centering
    \subfloat[ \label{fig:anglevalidation} ]{{\includegraphics[width=9.0cm, height=6.1cm]{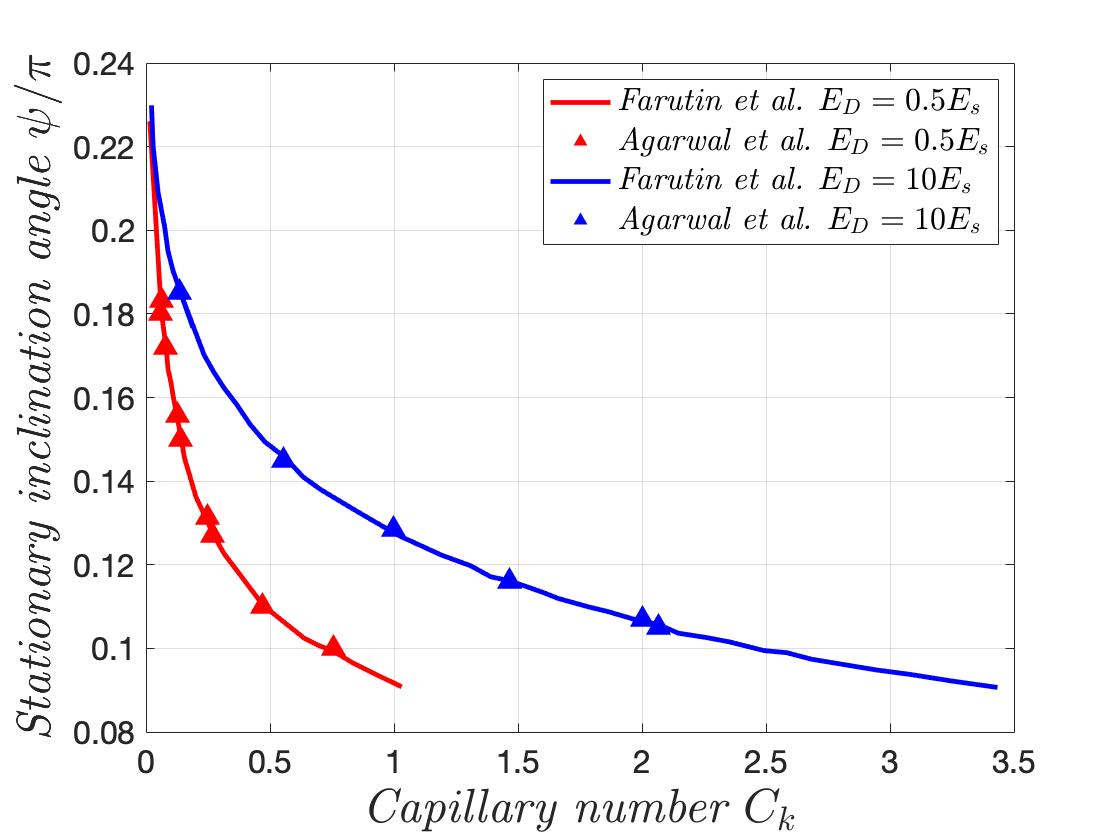} }}
    \subfloat[  \label{fig:capsule_validation} ]{{\includegraphics[width=9.0cm, height=6.1cm]{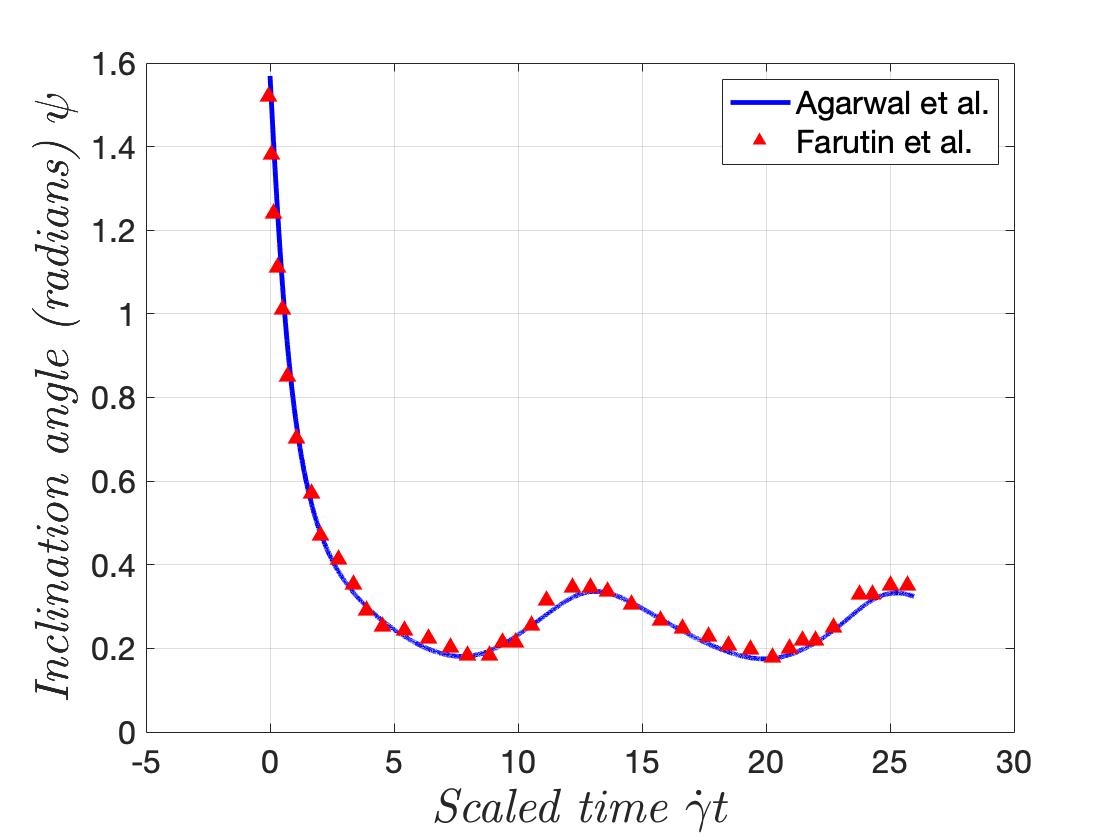} }}
    \caption{Comparison with known results from literature. (a) Stationary inclination angle of initially spherical extensible capsule simulated with Skalak model for shear elasticity. $\lambda =1, \kappa_{b} = 0.$ The solid lines are results from \cite{11} while the triangle and square symbols are our results. (b) Time evolution of inclination angle of inextensible capsule in unconfined shear flow for $C_{a} = 1, \lambda = 1, \nu = 0.65, \chi = 0.18$. Blue curve is our simulations while the red crosses are data points  extracted from Fig. 15 right of \cite{11} using WebPlotDigitizer software. We observe swinging-tank treading dynamics. We take the initial shape to be stress-free oblate and use the finite extensibility non-linear elasticity model for shear elastic energy proposed in \cite{11}. }%
    \label{fig:validation}%
\end{figure*}

In this section, we reproduce existing benchmark results to validate our model and the code. For this purpose, we simulate an initially spherical extensible capsule (without inextensibility constraint) in unconfined shear flow with viscosity contrast $\lambda = 1$ and bending modulus $\kappa_{b} = 0$. The volume equivalent radius $R_{0} = 1$ in these simulations. We calculate terminal inclination angles, $\psi$ (defined as the angle between the major axis of the capsule and the direction of the shear flow), for different capillary numbers. We summarize these results in in Fig. \ref{fig:anglevalidation} and compare it with the results reported in \cite{12, 11}. We observe good quantitative agreement.  

In a second validation computation, we reproduce tank treading with swinging dynamics of inextensible capsule in unconfined shear flow at $C_{a} = 1$ and plot the time evolution of the inclination angle in Fig. \ref{fig:capsule_validation}. We compare with the results reported in \cite{11} for the same parameters and observe good quantitative agreement. 

\subsection{\label{sec:shearflow} Shear flow }
In this section, we reproduce the existing experimental and numerical results on the shape dynamics of a red blood cell in shear flow regime. Our results are in good agreement with experiments demonstrating the physical relevance of our simulations. We also view this as an additional validation of our code. 

\begin{figure}%
    \centering
    \subfloat[]{{\includegraphics[width=3.36cm, height=2.66cm]{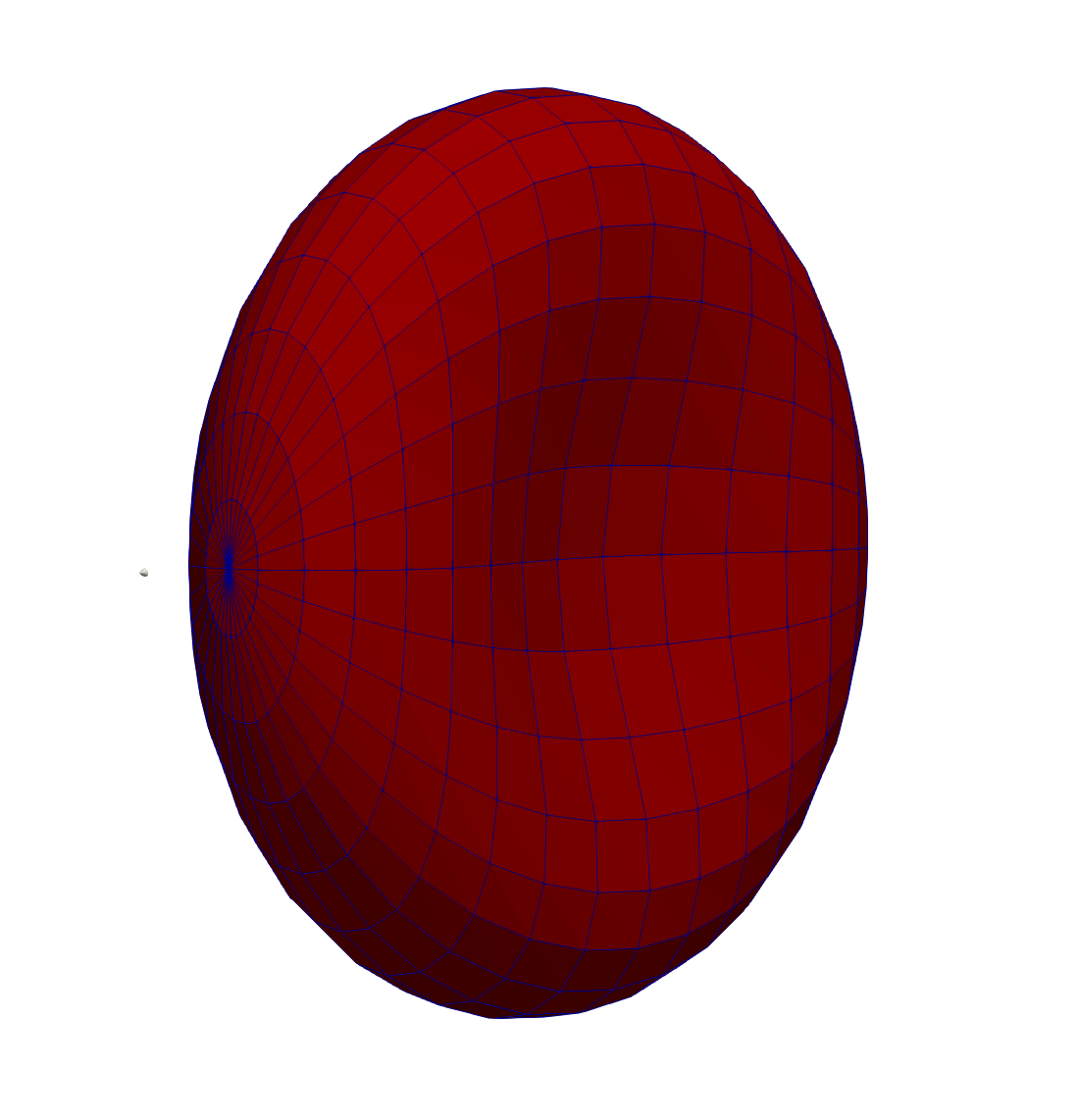} }}
    \subfloat[   ]{{\includegraphics[width=3.36cm, height=2.66cm]{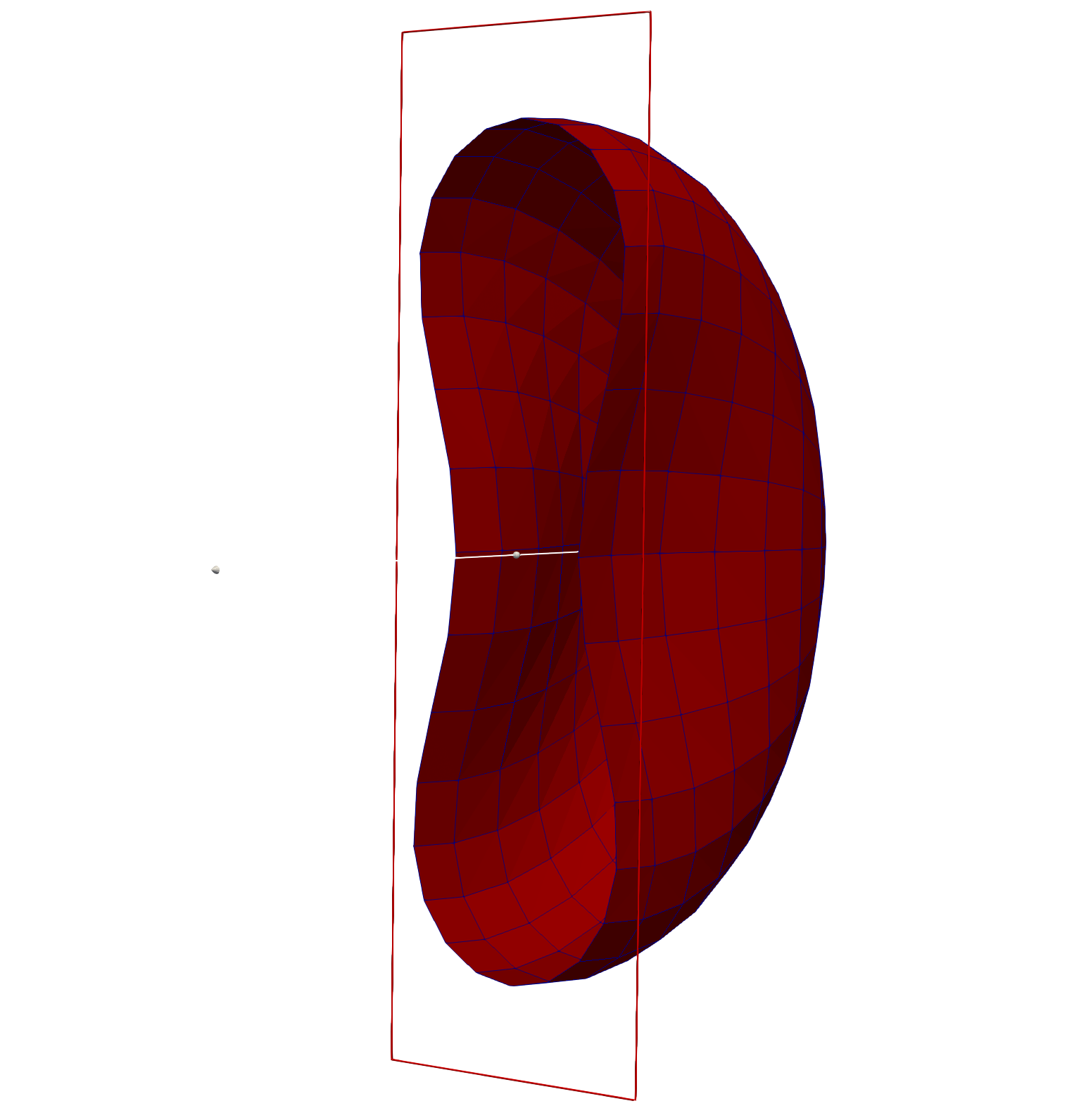} }}
   
   \caption{ (a) Initial biconcave shape of the capsule. (b)  Cross sectional view of the initial shape. }
   \label{fig:biconcave}
\end{figure}

\begin{figure*}
\includegraphics[width=9cm]{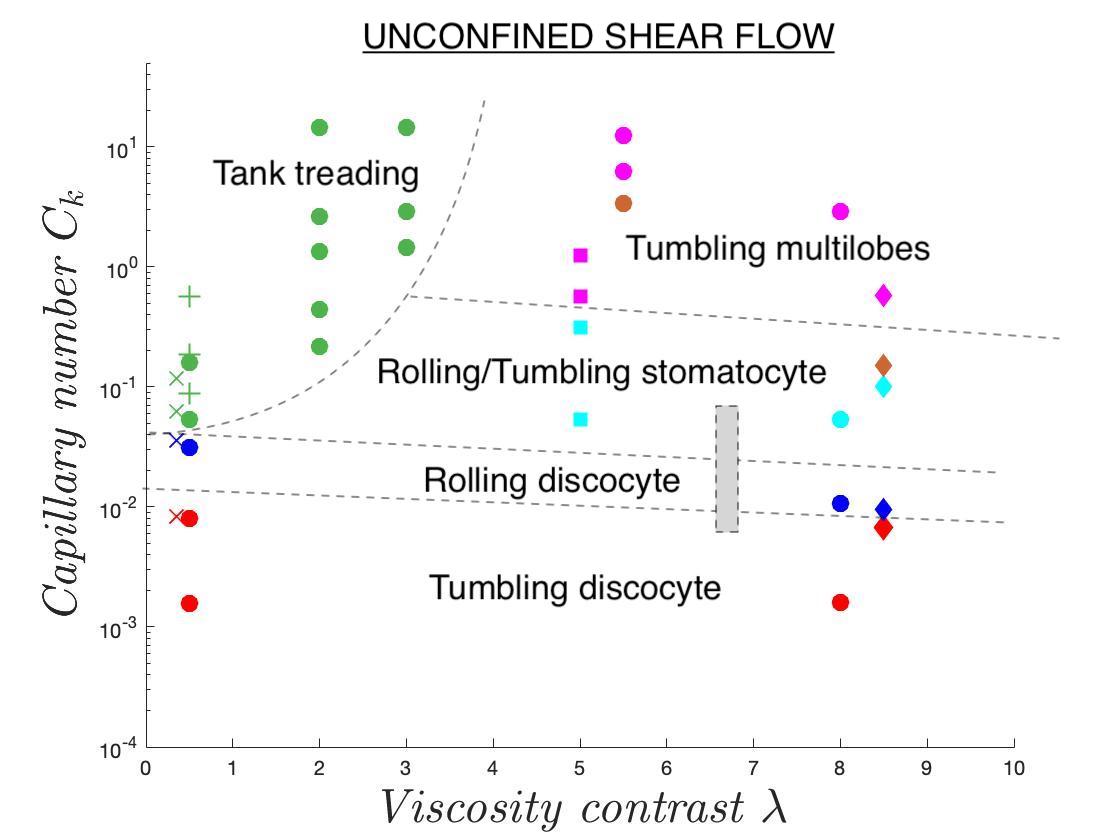}
\caption{Phase diagram for shear flow dynamics of RBC in the parameter space of capillary number $C_{k}$ and viscosity contrast $\lambda$. Regions corresponding to different shapes are separated by a dashed line . Red denotes tumbling discocyte state, green denotes tank treading, blue denotes rolling discocyte,  cyan denotes rolling stomatocytes, brown denotes tumbling stomatocytes and magenta denotes tumbling multilobes. Circles are our simulations,  $\times$ are the experimental results in \cite{20}, $+$ are the experimental results from \cite{24}, diamonds are the experimental results from \cite{23},  squares are the experimental results from \cite{26}. The grey shaded block represents the range of tumbling to rolling transition observed in \cite{25}. Experimental capillary numbers are obtained using $R_{0} = 3.25 \times 10^{-6}m $ and $E_{s} = 4.8 \times 10^{-6} kg \cdot s^{-2}$.  Experimental data points are the most frequent states observed at those parameters. Rolling behavior in simulations is observed when capsule's axis is out of the shear plane initially. }
\label{fig:phasecapshear}
\end{figure*}

\begin{figure}%
    \centering
    \subfloat[]{{\includegraphics[width=3.15cm, height=2.45cm]{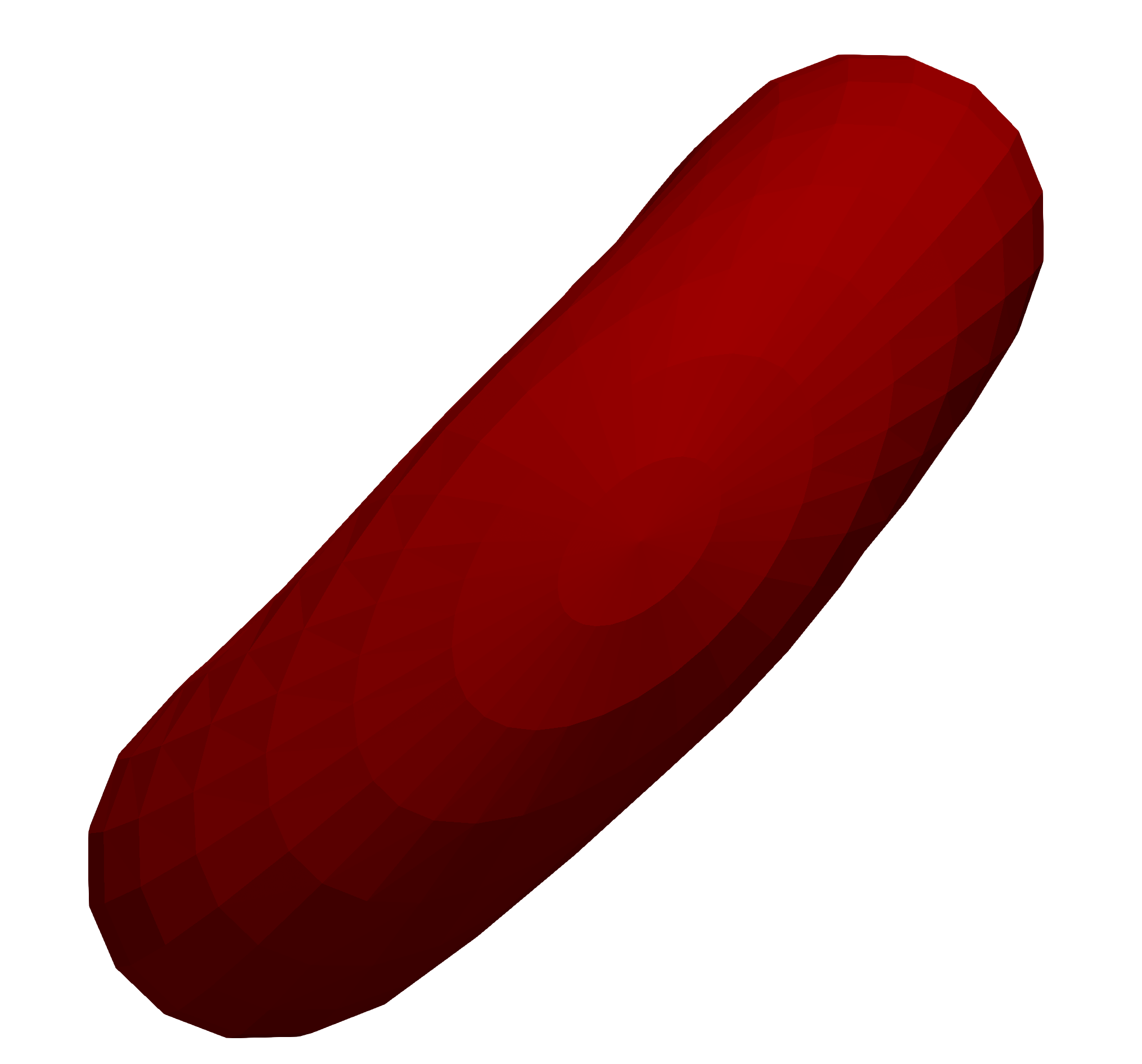} }}
    \subfloat[   ]{{\includegraphics[width=3.36cm, height=2.66cm]{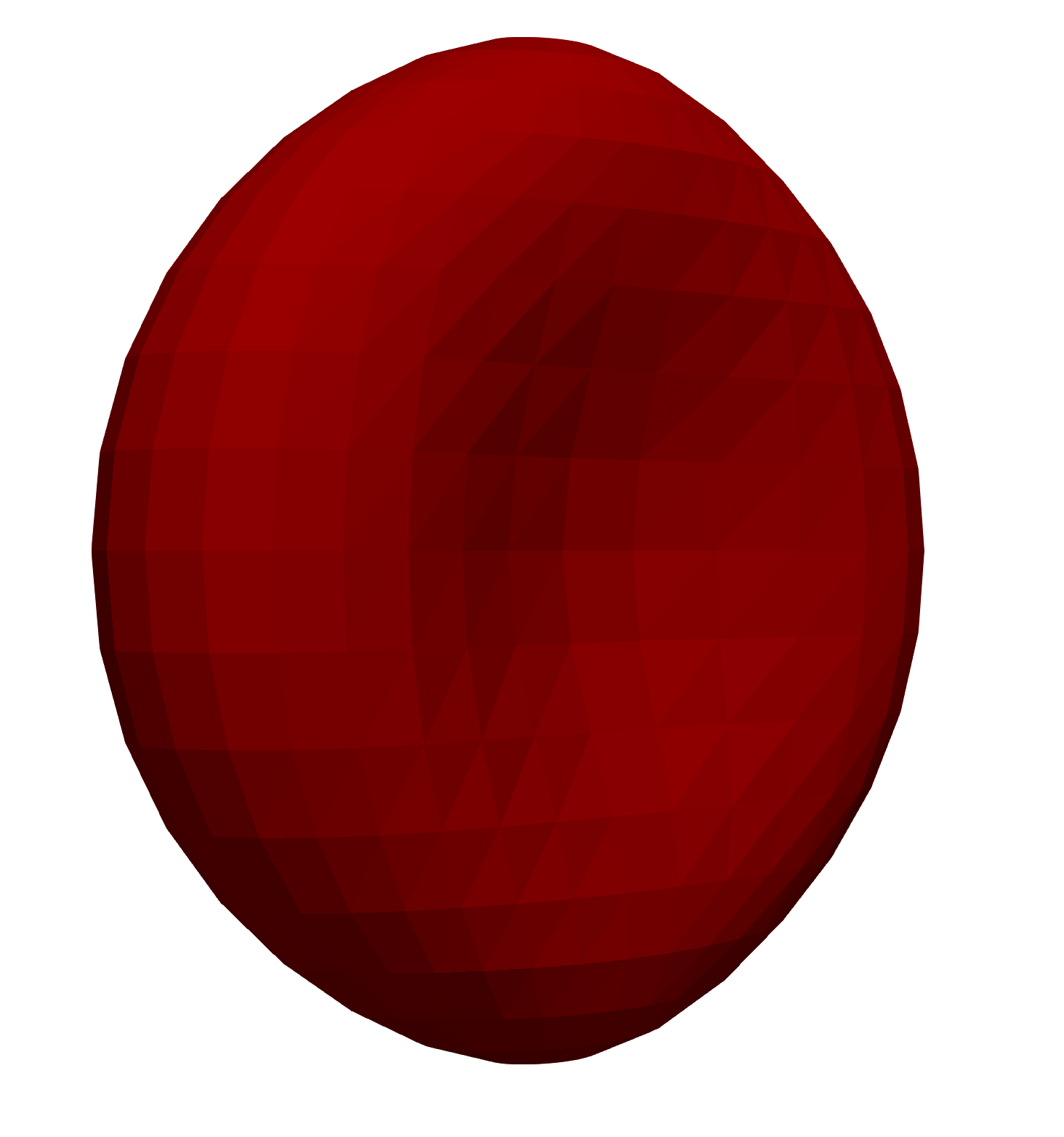} }}
    \caption{Tumbling discocyte for $\lambda = 0.5, C_{k} = 0.001$.  (a) Side view. (b) Top view. \label{fig:tumbdisc} }%
    \subfloat[]{{\includegraphics[width=3.15cm, height=2.52cm]{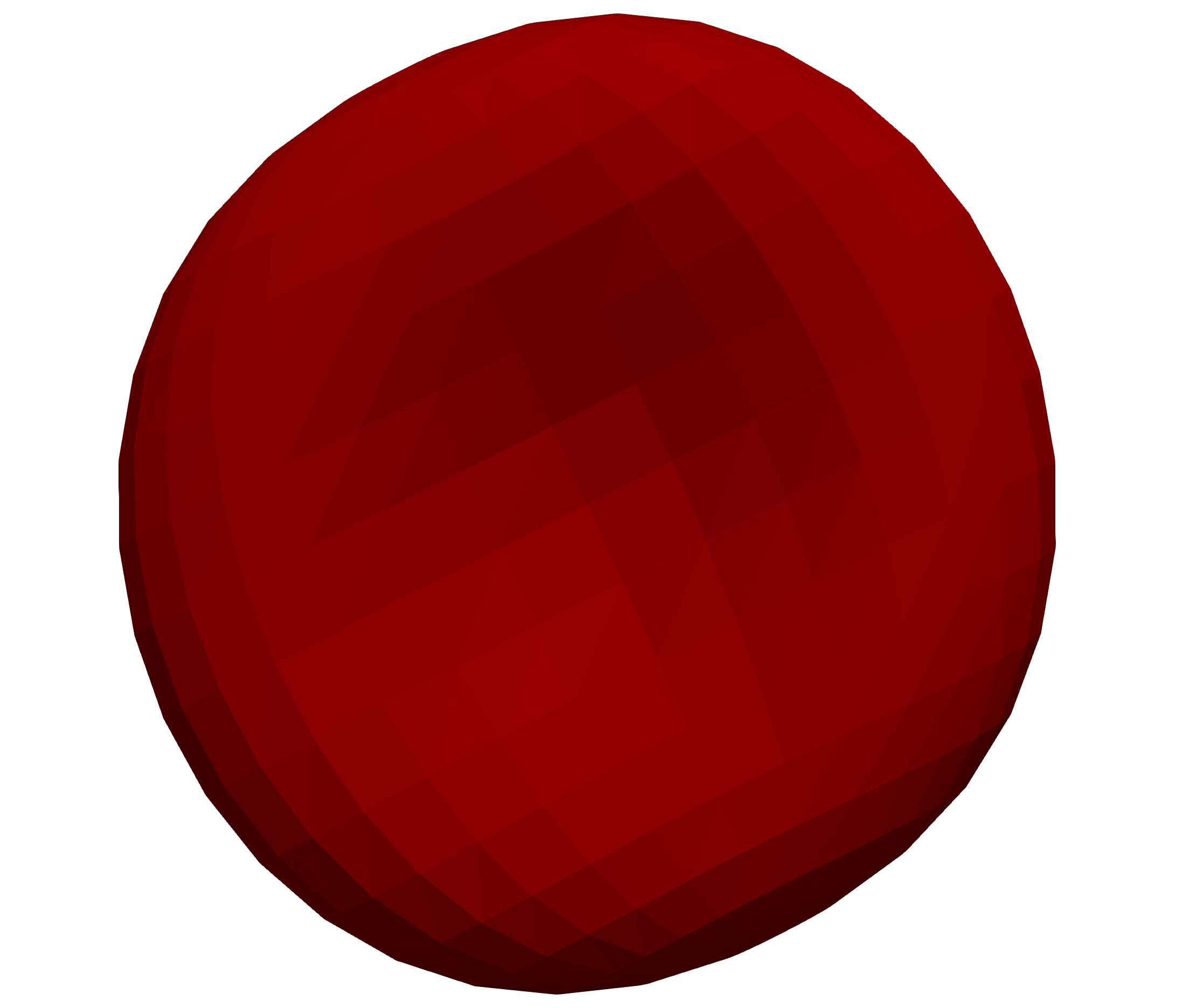} }}
    \subfloat[  ]{{\includegraphics[width=3.64cm, height=3.08cm]{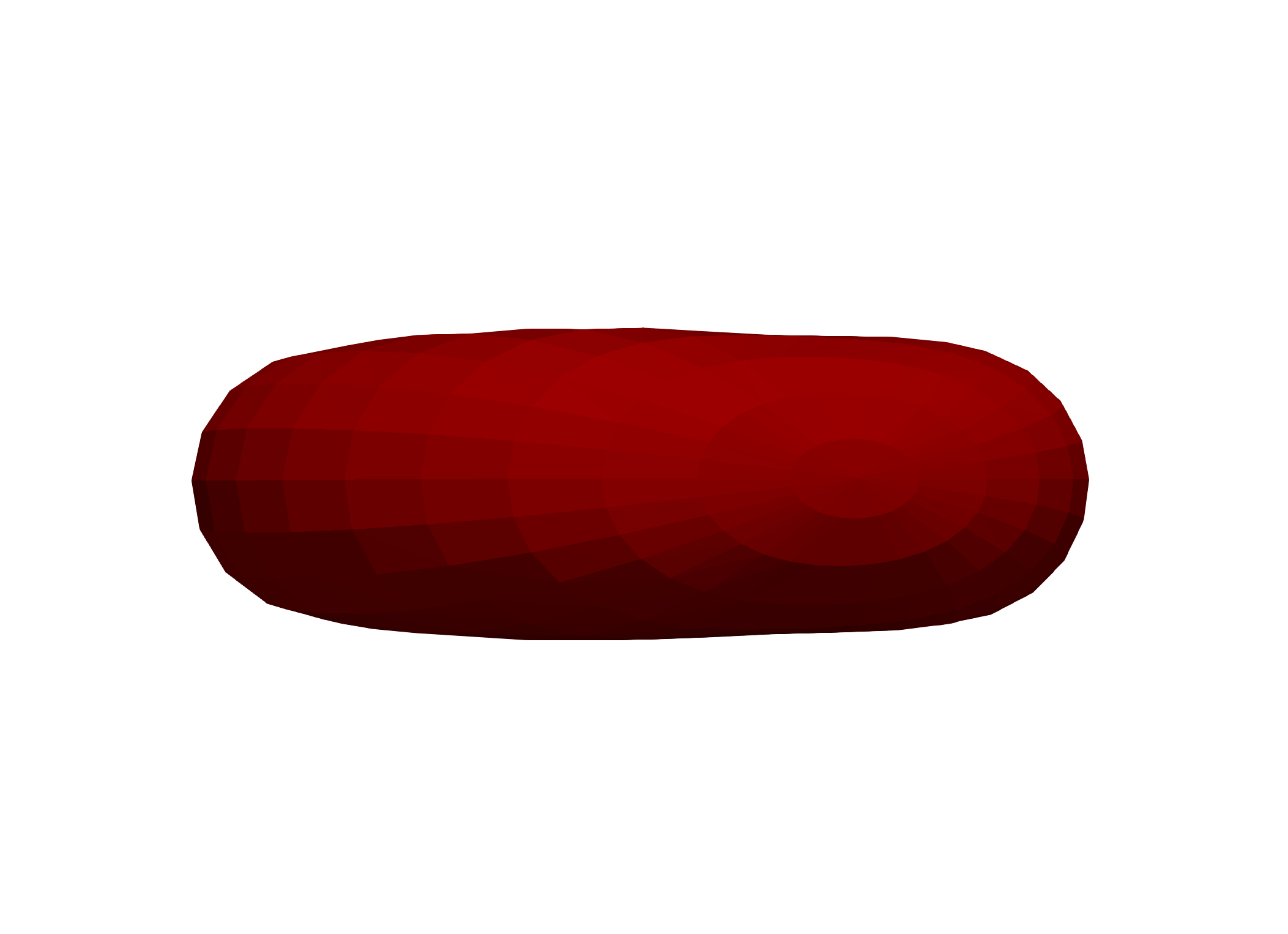} }}
    \caption{Rolling discocyte for $\lambda = 0.5, C_{k} = 0.02$. (a) Side view. (b) Top view. \label{fig:rolldisc} }%
    \subfloat[]{{\includegraphics[width=3.15cm, height=2.52cm]{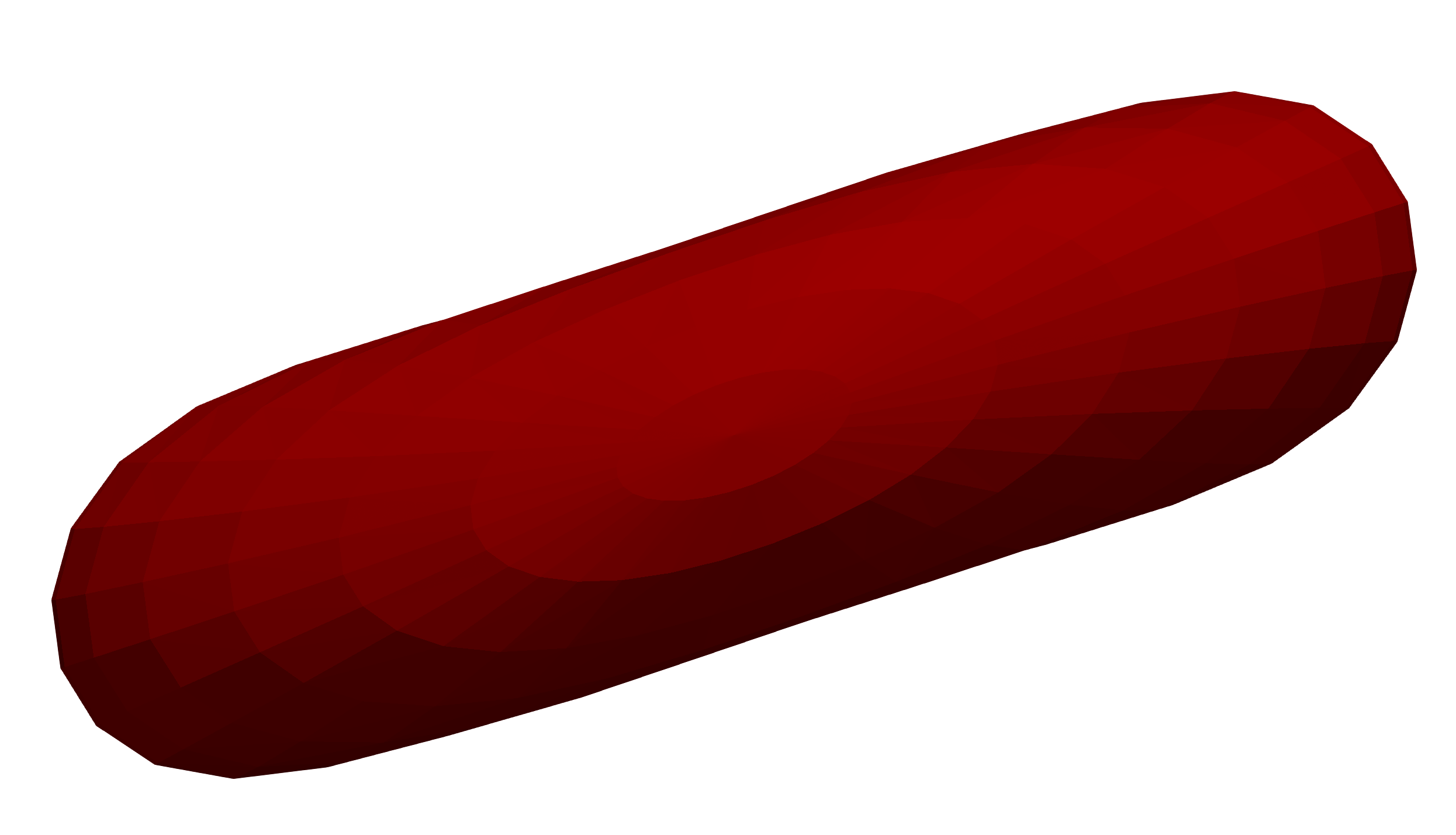} }}
    \subfloat[  ]{{\includegraphics[width=3.36cm, height=2.66cm]{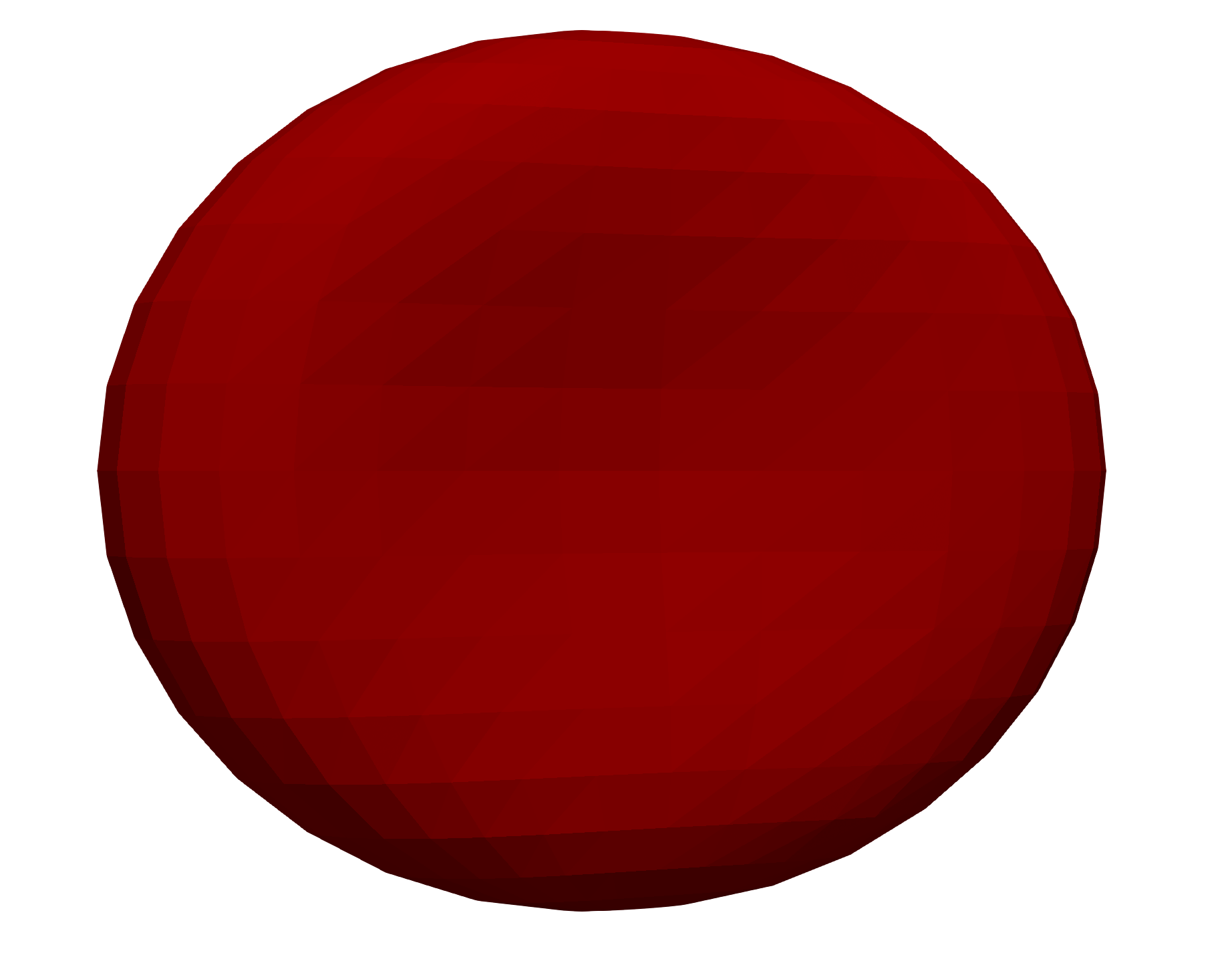} }}
    \caption{ Tank treading discocyte for $\lambda = 0.5, C_{k} = 0.13$. (a) Side view. (b) Top view. \label{fig:ttdisc} }%
    \subfloat[]{{\includegraphics[width=3.15cm, height=1.862cm]{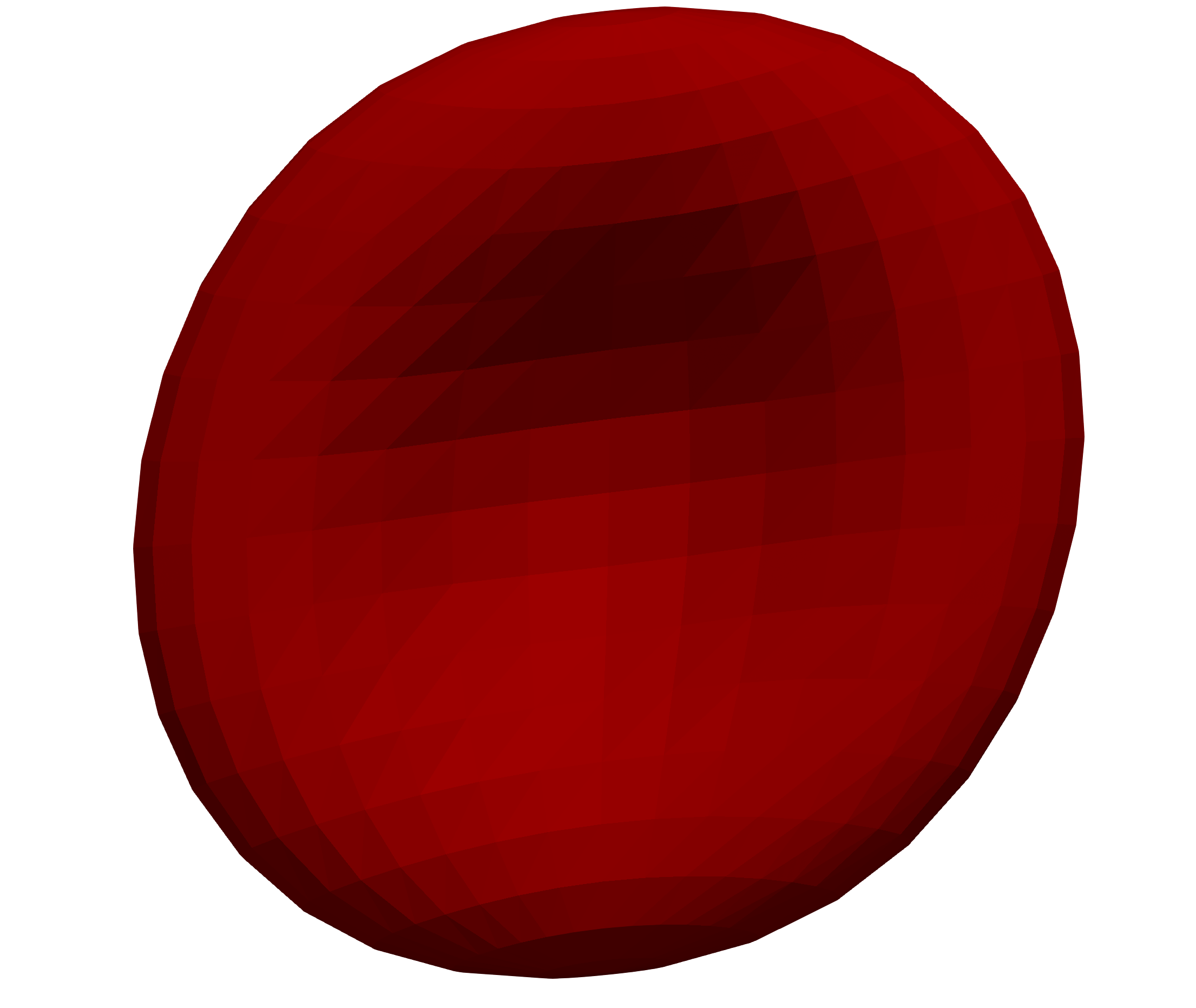} }}
    \subfloat[  ]{{\includegraphics[width=3.36cm, height=2.56cm]{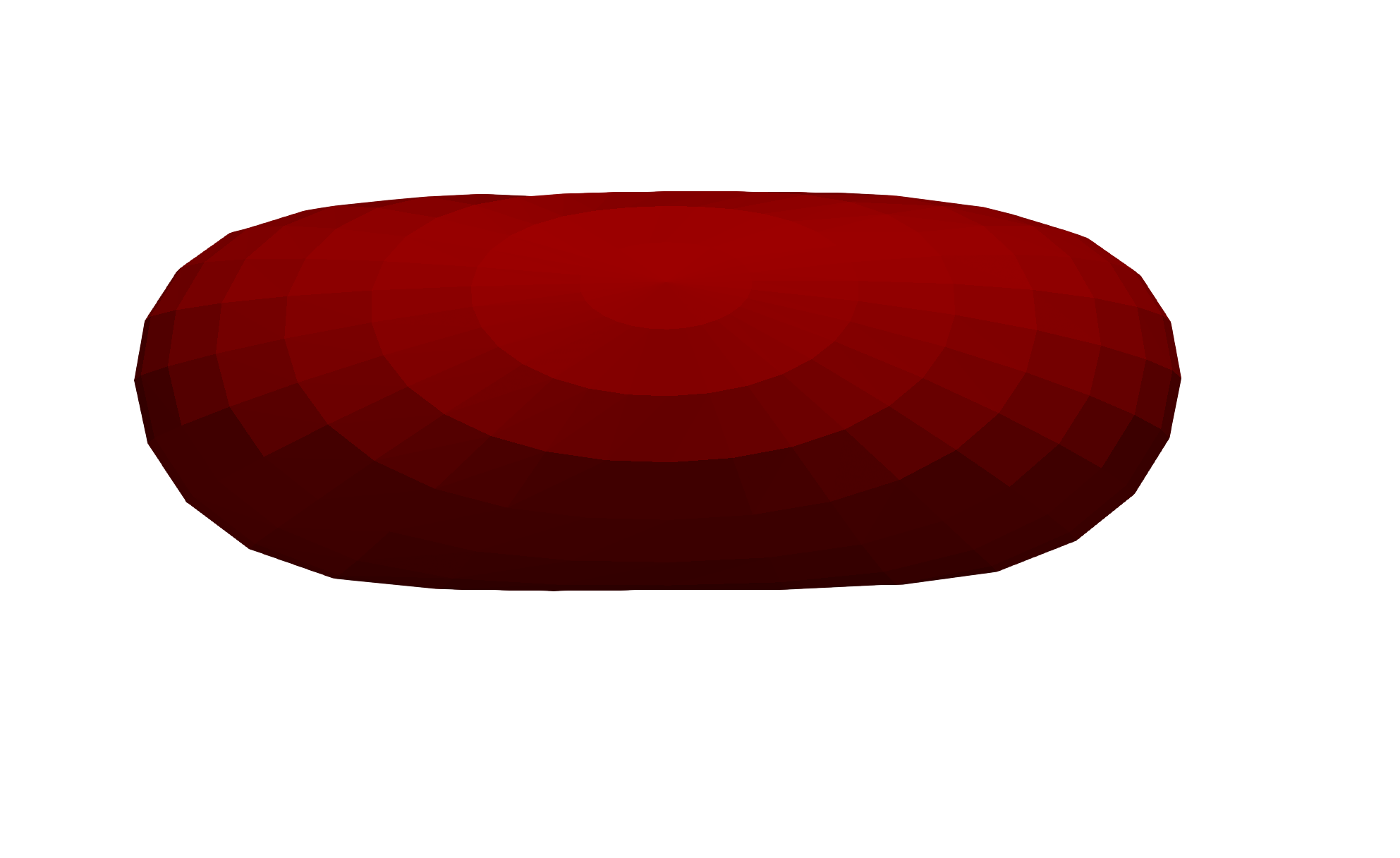} }}
    \caption{Rolling stomatocyte for $\lambda = 8, C_{k} = 0.045$. (a) Side view. (b) Top view. \label{fig:rollstom} }%
     \subfloat[]{{\includegraphics[width=3cm]{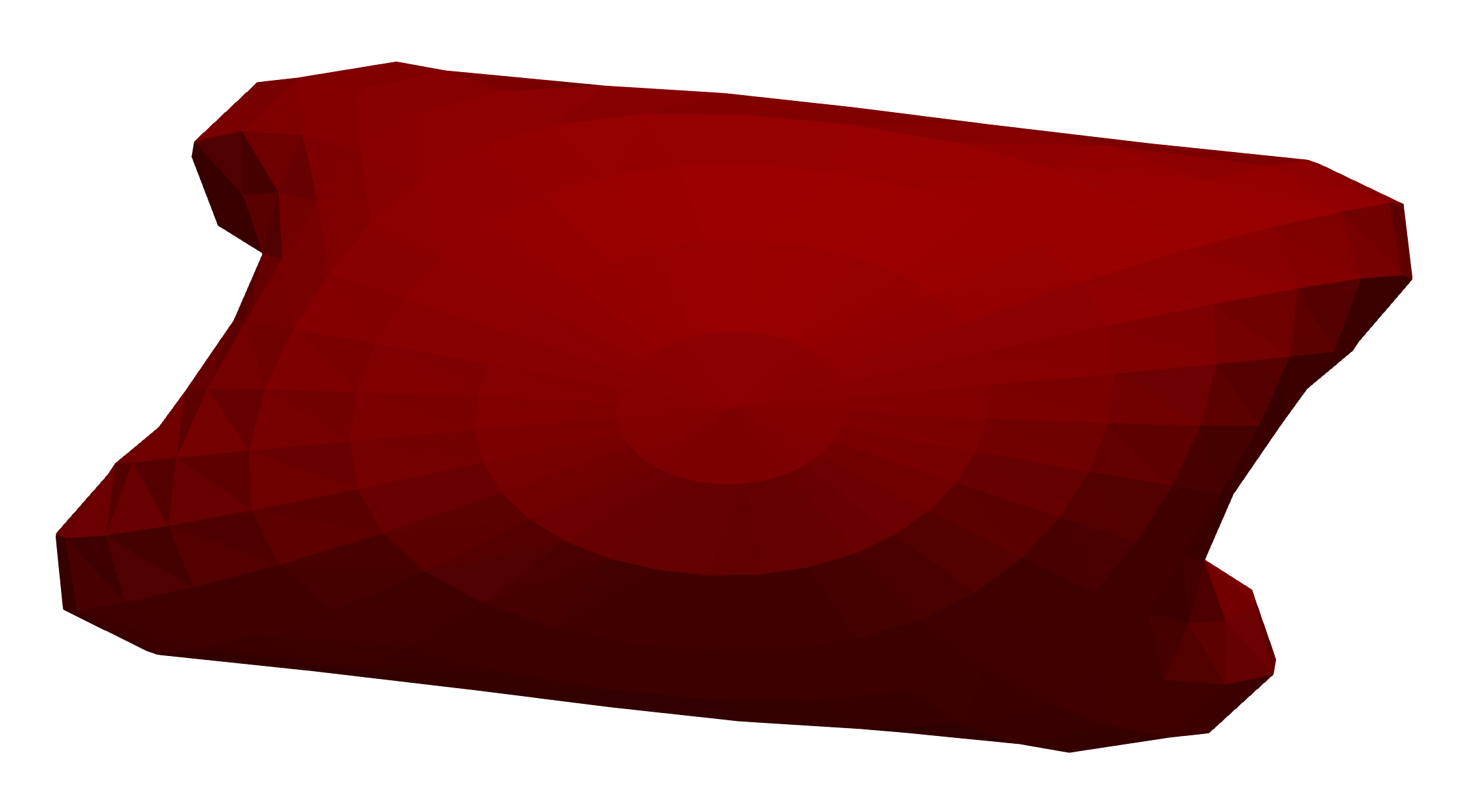} }}
    \subfloat[  ]{{\includegraphics[width=3.36cm, height=2.66cm]{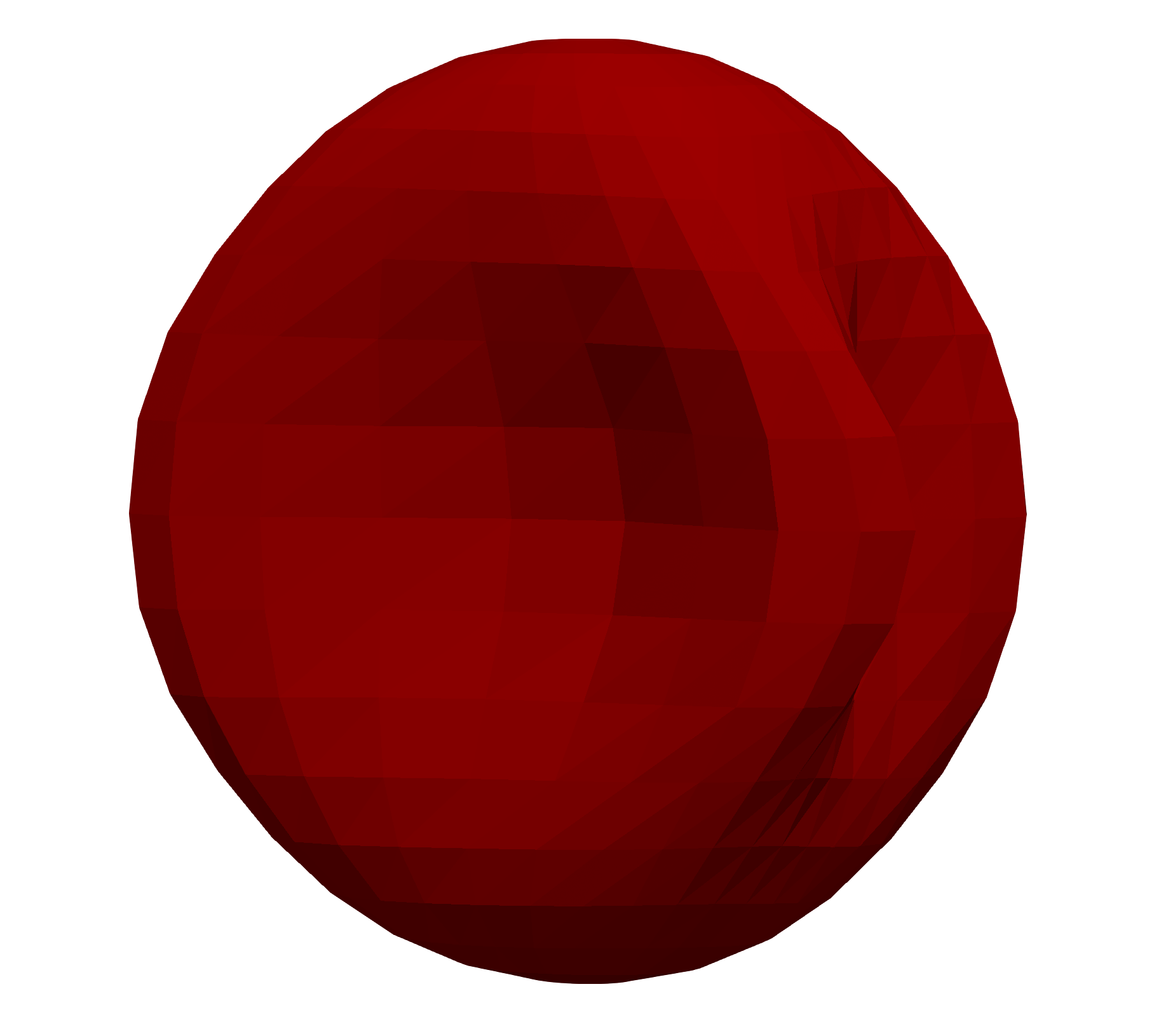} }}
    \caption{ Tumbling multilobes for $\lambda = 5, C_{k} = 5.2$. (a) Side view. (b) Top view. \label{fig:tumblobe}}%
   
\end{figure}

Early experimental studies \cite{2, 13, 3} demonstrated that a RBC in linear shear flow shows a transition from tumbling (TB) dynamics (similar to that observed in rigid bodies) to tank treading (TT) dynamics as the shear-force capillary number is increased. The tank treading motion is also accompanied by oscillations in the inclination angle, which is called swinging (SW)\cite{14}. Numerical simulations with initially oblate capsules \cite{15, 16, 11} have also demonstrated the transition from TB to TT-SW dynamics as the shear-force capillary number is increased. Recent experiments \cite{20, 23, 24, 25, 26} have shown the existence of rolling dynamics at intermediate capillary numbers between the TB and TT regime. In particular, the experiments in \cite{23, 26} showed that higher viscosity contrasts ($\lambda > 3$) lead to stomatocytes and multilobe shapes instead of the TT dynamics observed for $\lambda \leq 3$ as the shear-force capillary number is increased. Initial numerical studies could not reproduce the TB-TT transition as a function of the flow shear rate; for this reason researchers introduced residual stresses \cite{17,18,19} by considering that the stress-free configuration of the capsule corresponds to a near sphere. Following these works, we assume that the reference configuration is an oblate spherical capsule of reduced volume 0.96. We deflate it to an oblate capsule of reduced volume 0.65 with $R_{0} = 3.25$ and let it relax to a biconcave shape (in dimensional terms it corresponds to area $A = 134 \mu m^{2}$ and volume $V = 94 \mu m^{2}$, which are consistent with experimentally obtained values \cite{18, 26}; see Fig. \ref{fig:biconcave}). The bending stiffness ratio is chosen to be $\chi = 0.025$. Our dimensionless parameters correspond to shear modulus $E_{s} = 1.25 \times 10^{-6} kg \cdot s^{-2}$ and bending modulus $\kappa_{b} = 2.5 \times 10^{-19} J$. In this section, we reproduce the experimental shear flow results using our code. We observed good quantitative agreement with the experimental results. We summarize the results in the phase diagram in Fig. \ref{fig:phasecapshear}. We confirm previous observations in which at low viscosity contrast ($\lambda \leq 3$), tumbling (see Fig. \ref{fig:tumbdisc}) to tank treading (see Fig. \ref{fig:ttdisc}) transition takes place as the shear-force capillary number is increased. We also observe the rolling dynamics (see Fig. \ref{fig:rolldisc}) at intermediate capillary numbers between the TB and TT regime.  At higher viscosity contrasts ($\lambda > 3$), no tank treading is observed; rather we observe stomatocyte shape (see Fig. \ref{fig:rollstom}) which then transitions to tumbling lobar shapes (see Fig. \ref{fig:tumblobe}) as $C_{k}$ is increased further. 

\begin{figure}%
    \centering
    \subfloat[ ]{{\includegraphics[width=3.5cm, height=2.4cm ]{cap1405f_paper.png} }}
    \subfloat[ ]{{\includegraphics[width=3.0cm, height=2.6cm]{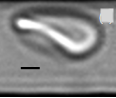} }}
    \subfloat[ ]{{\includegraphics[width=3.5cm,  height=2.4cm]{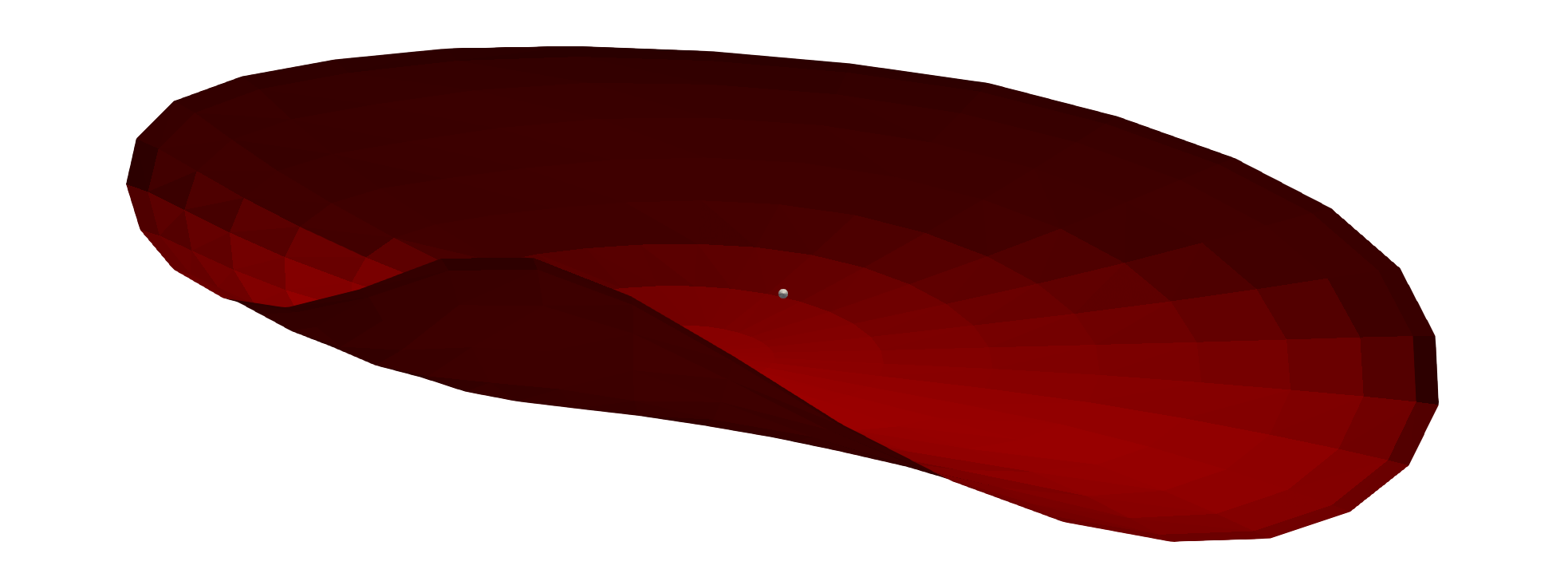} }}
    \caption{Slipper shape observed in simulations and experiments. (a) Side view of slipper in our simulations of confined flow with $\lambda=5, C_{n} = 0.55, C_{k} = 0.4$. (b) Cross sectional view of slipper in our simulations for same parameters as in (a). (c) Side view of slipper in experiments in \cite{41} for $ \lambda=5, C_{n} \approx 0.54, C_{k} = 0.35$.  }%
    \label{fig:slipper}%
\end{figure}

\begin{figure}%
    \centering
    \subfloat[ ]{{\includegraphics[width=3.3cm, height=3.2cm]{cap1338f_paper.png} }}
    \subfloat[ ]{{\includegraphics[width=3.0cm, height=2.6cm]{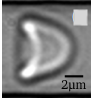} }}
    \subfloat[ ]{{\includegraphics[width=2.8cm,  height=2.7cm]{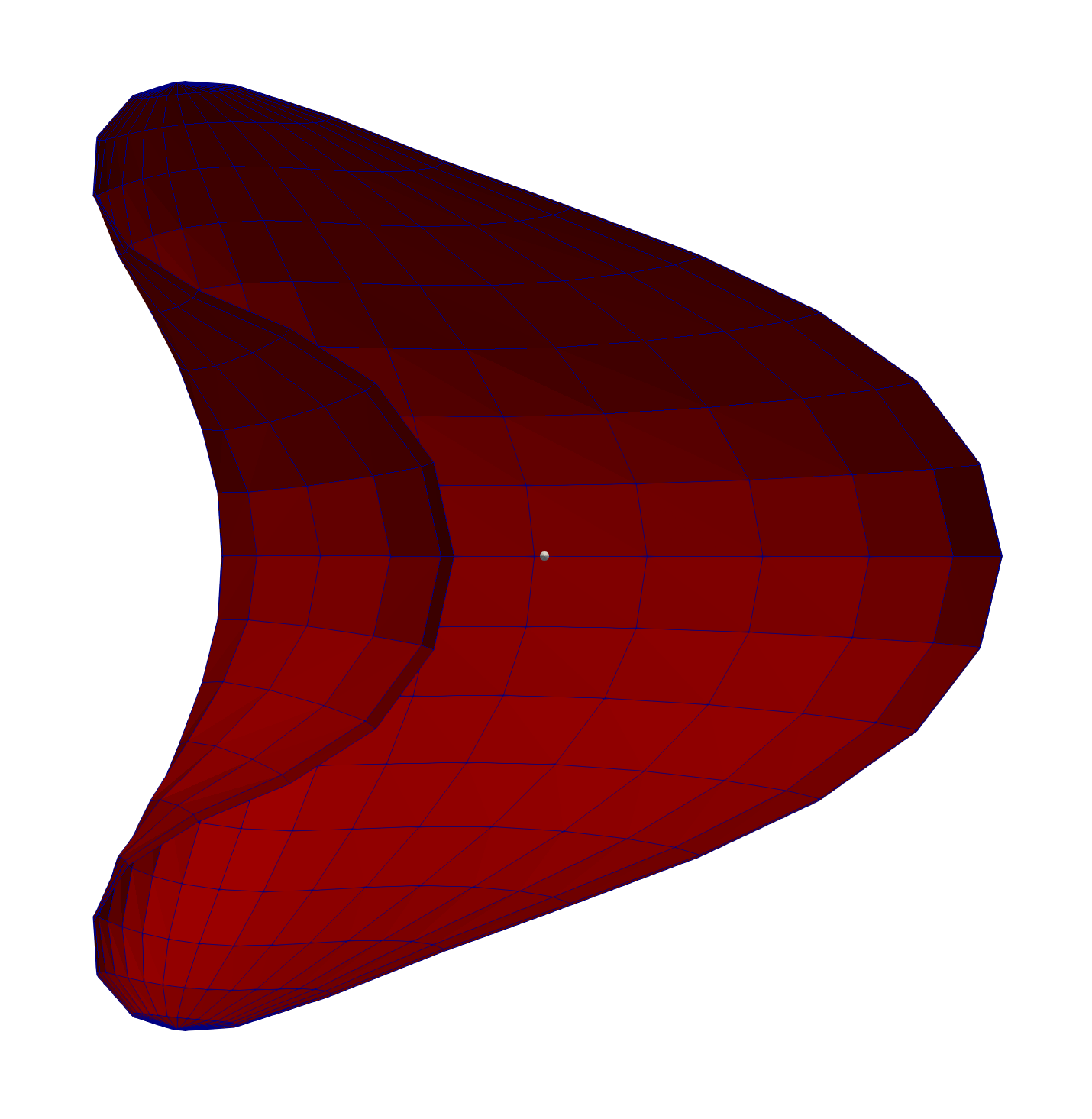} }}
    \caption{Croissant shape observed in simulations and experiments. (a) Side view of croissant in our simulations of confined flow with $\lambda=5, C_{n} = 0.55, C_{k} = 0.1$. (b) Cross sectional view of croissant in our simulations for same parameters as in (a). (c) Side view of croissant in experiments in \cite{41} for $ \lambda=5, C_{n} \approx 0.54, C_{k} \approx 0.07$.  }%
    \label{fig:croissant}%
\end{figure}

\section{\label{sec:results} Results}
In this section, we discuss the shape dynamics of a capsule in unconfined and confined Poiseuille flow ($0.25 \leq C_{n} \leq 0.65$) for both $\lambda=1$ and $\lambda=5$. The capsule membrane parameters, the initial shape and the residual stress is the same as used in the shear flow simulations. 

\subsection{\label{sec:uncon1poiseuilleflow} Unconfined Poiseuille flow $\lambda =1$}
\begin{figure*}%
    \centering
    \subfloat[Viscosity contrast $\lambda=1$ \label{fig:phasecapuncon1} ]{{\includegraphics[width=9cm]{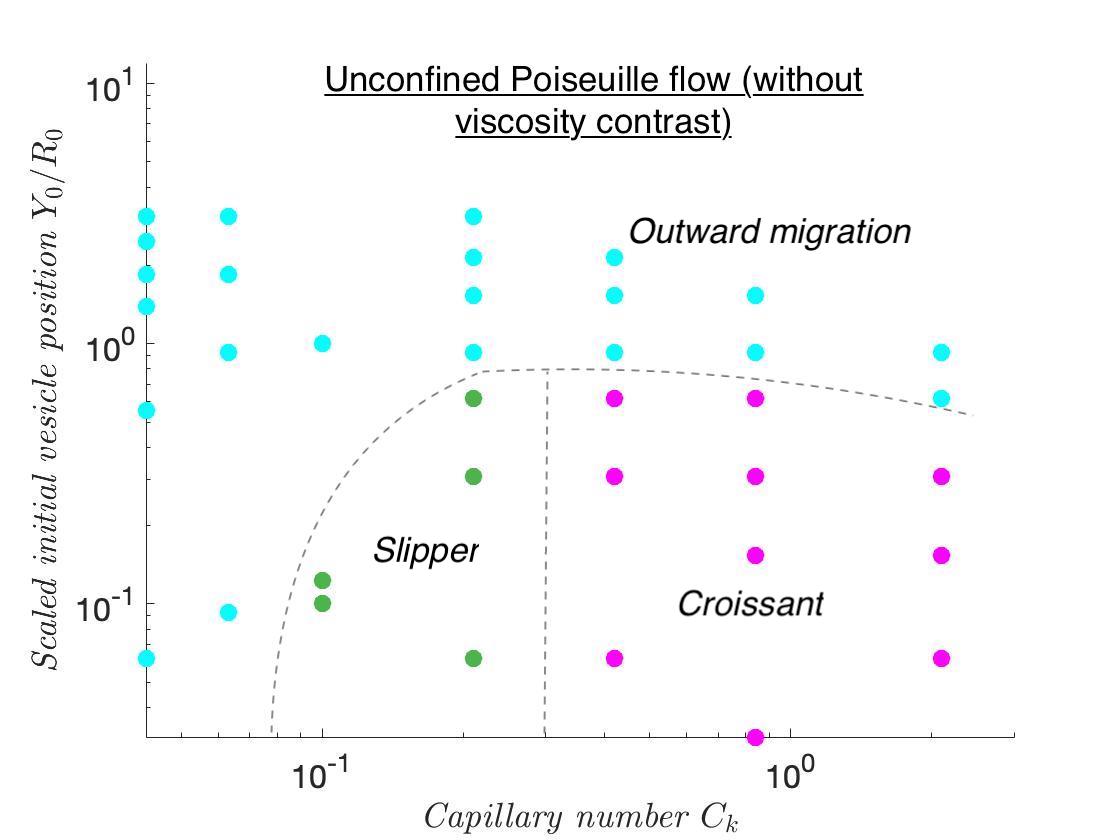} }}
    \subfloat[ Viscosity contrast $\lambda=5$  \label{fig:phasecapuncon5} ]{{\includegraphics[width=9cm]{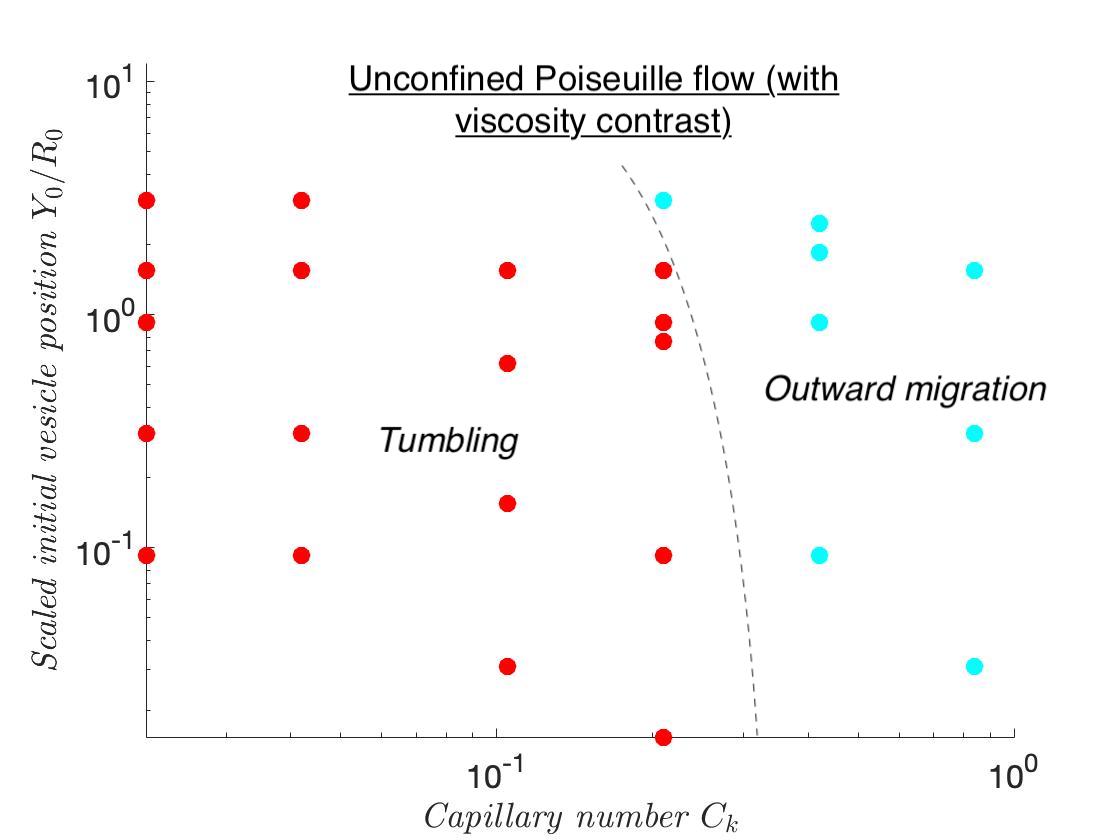} }}
   
   \caption{(a) Phase diagram for unconfined Poiseuille flow dynamics of RBC in the parameter space of capillary number $C_{k}$ vs vesicle scaled initial position $Y_{0}/R_{0}$ for $\lambda = 1$.  (b)  Phase diagram for unconfined Poiseuille flow dynamics of RBC in the parameter space of capillary number $C_{k}$ vs vesicle scaled initial position $Y_{0}/R_{0}$ for $\lambda = 5$. Regions corresponding to different shapes are separated by a dashed line. Cyan denotes outward migration, green denotes tank treading slipper, magenta denotes croissant/parachute shape and red denotes tumbling. Circles indicate our simulations.The dashed lines serve as a guide to the eye.}
\end{figure*}

We study the shape dynamics of a capsule for varying capillary numbers and initial positions ($Y_{0}$) from the centerline of the flow for $\lambda=1$. We observe three different dynamics, namely, \textit{outward migration}, \textit{tank treading slipper} and \textit{croissant }.  shape. The slipper and croissant shapes are similar to the ones shown in Fig. \ref{fig:slipper} and Fig. \ref{fig:croissant} respectively. For low capillary numbers $0.04 \leq C_{k} \leq 0.07$, we observe an outward migration with a slipper-like shape at all initial positions in our simulations ($Y_{0}/R_{0} \geq 0.015$). For $C_{k} =0.21 $, we observe a coexistence of slipper and outward migration dynamics, i.e.,  TT slippers for $0.015 \leq Y_{0}/R_{0} \leq 0.61$ and outward migration for $Y_{0}/R_{0} > 0.61$. As the capillary number is increased higher to $C_{k} \geq 0.42$, we observe croissants for lower initial positions and outward migration for higher initial positions. We combine all these results to plot a phase diagram in the parameter space of capillary number $C_{k}$ and the capsule's scaled initial position $Y_{0}/R_{0}$ shown in Fig. \ref{fig:phasecapuncon1}. 

\textit{Discussion of results:} The transition from slipper to croissant with increase in capillary number is similar to the 3D vesicle behavior as observed in previous numerical studies \cite{22}.  The stable slipper position with respect to the centerline of the flow decreases as the capillary number increases like observed with 3D vesicles in \cite{22}. However, the outward migration results observed for capsules are strikingly different from the results on 3D vesicles. The numerical studies on 3D vesicles \cite{22, 37} have shown that for $\lambda=1$ (no viscosity contrast) no outward  migration is observed.  For capsules, we observe outward migration for high initial positions even when no viscosity contrast is present. The observation of outward migration regime at high initial positions extends the conclusion of our study on 3D vesicles \cite{22} that the outward migration tendency, when exists, increases with increase in initial position. Another surprising result of this study is that the outward migration seems to be the only dynamics present at low capillary numbers ($C_{k} \leq 0.08$) while both slipper and outward migration are observed at intermediate $C_{k}$. This could be possibly be explained due to our observation that stable slipper positions rise with decrease in capillary number $C_{k}$. As the capillary number $C_{k}$ decreases, the increase in slipper position at stable configuration leads to increase in outward migration tendency. At some critical capillary number (critical $C_{k}$ value about $0.08$ here), the outward migration could dominate causing the slipper to become unstable and keep migrating outwards away from the centerline of the flow.     
\subsection{\label{sec:uncon5poiseuilleflow} Unconfined Poiseuille flow $\lambda =5$}
For a capsule in unconfined Poiseuille flow with $\lambda=5$, we observe two regimes, namely, \textit{tumbling} and \textit{outward migration}. For low capillary numbers $C_{k} < 0.21$, we observe tumbling  dynamics at all initial capsule positions. As the capillary number is increased to $C_{k} = 0.21 $, we observe tumbling for initial positions $Y_{0}/R_{0} \leq 1.8$ and outward migration for $Y_{0}/R_{0} > 1.8$. For even higher capillary numbers $C_{k} > 0.35$, at all initial positions we see simultaneous migration and tumbling. The tumbling shapes vary from slipper to trilobes and multilobes depending on the local shear rate. We plot these results in phase diagram of capillary number $C_{k}$ vs scaled initial position $Y_{0}/R_{0}$ in Fig. \ref{fig:phasecapuncon5}. 

The tumbling dynamics (instead of the tank treading slipper as in $\lambda=1$) is consistent with the dynamics of capsules in shear flow (see Fig. \ref{fig:phasecapshear}). The shear flow dynamics show that for $\lambda=5$ there exists a tumbling multilobe/stomatocyte regime instead of the tank treading regime seen for $\lambda \leq 3$. However, no outward migration takes place for low capillary numbers (unlike for $\lambda=1$) and this is surprising.  This could be due to the tumbling dynamics observed for $\lambda=5$ instead of the tank treading slipper-like shapes for $\lambda=1$. This difference in dynamics seems to change the migration behavior significantly.    

 \subsection{\label{sec:conpoiseuilleflow} Confined Poiseuille flow $\lambda=1$ }
\begin{figure*}%
    \centering
    \subfloat[Viscosity contrast $\lambda=1$ \label{fig:phasecapcon1}]{{\includegraphics[width=9cm]{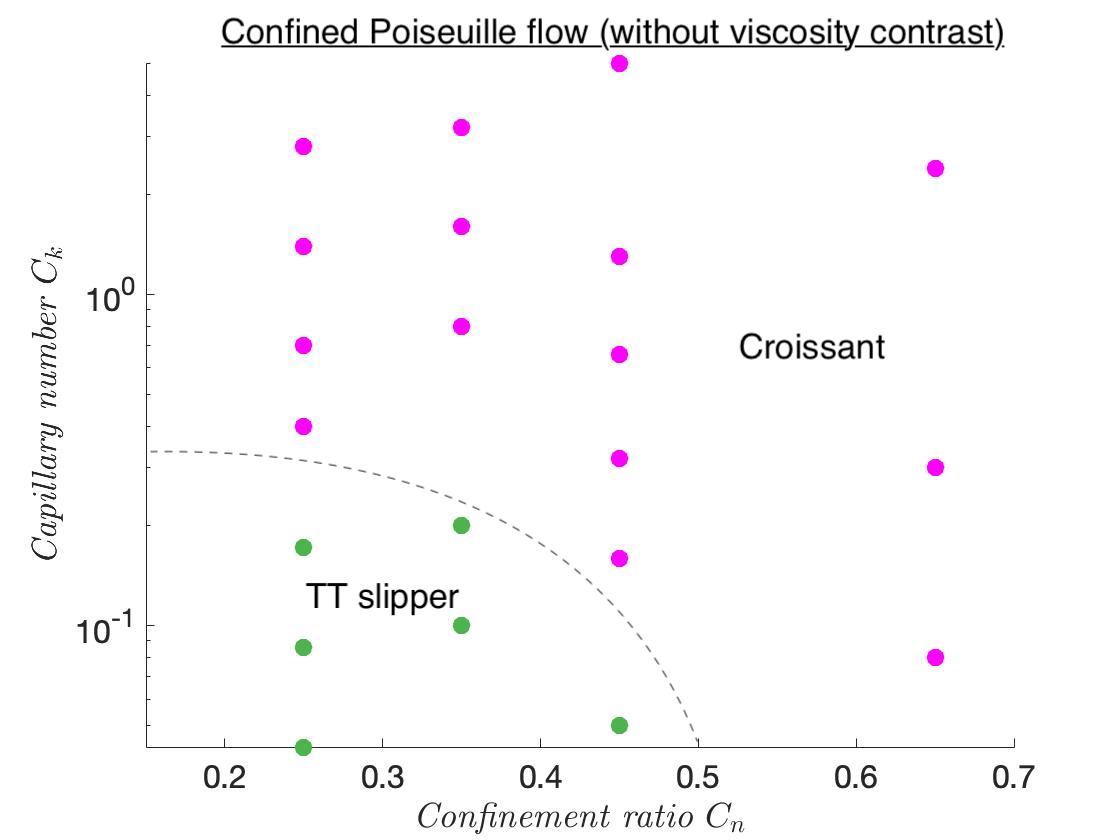} }}
    \subfloat[ Viscosity contrast $\lambda=5$ \label{fig:phasecapcon5} ]{{\includegraphics[width=9cm]{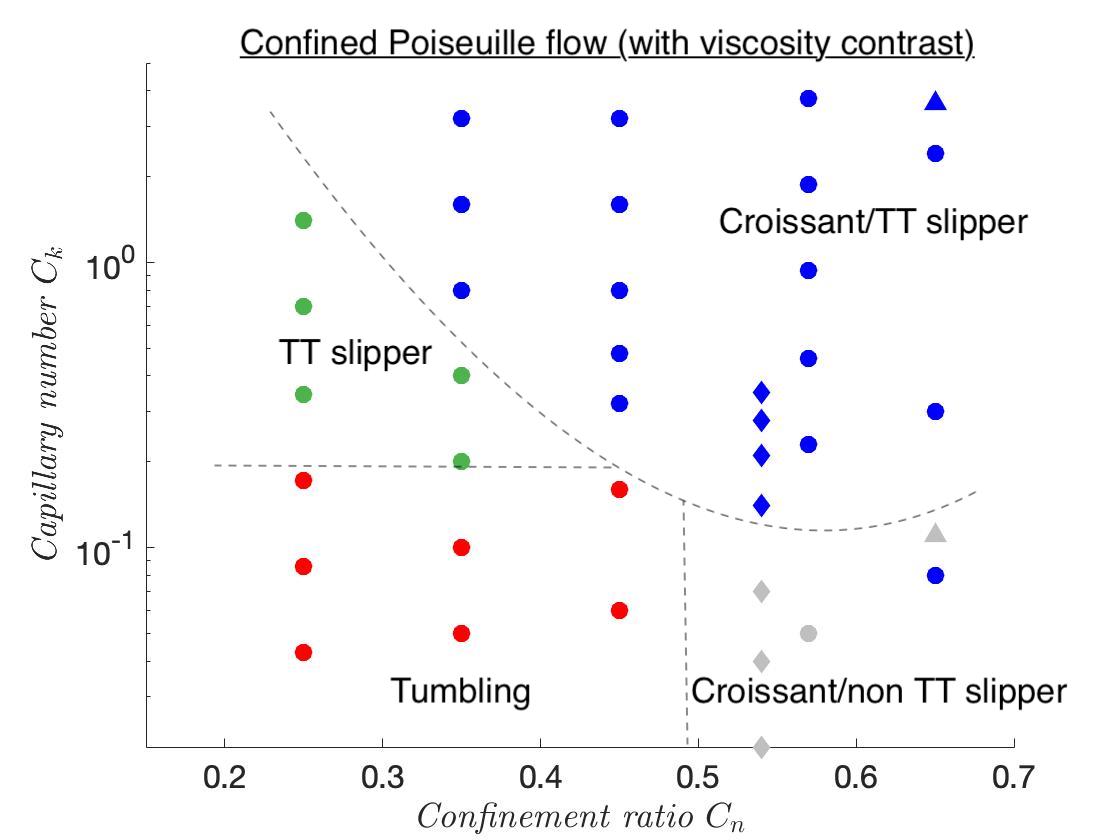} }}
   
   \caption{ Phase diagram for confined Poiseuille flow dynamics of RBC in the parameter space of confinement ratio $C_{n}$ vs capillary number $C_{k}$ for (a) $\lambda = 1$, and  (b) $\lambda=5$. Regions corresponding to different shapes are separated by a dashed line. Circles are our simulations, diamonds are the experimental results from \cite{41}, triangles are the experimental results in \cite{30}. Experimental capillary numbers are obtained using $R_{0} = 3.25 \times 10^{-6}m $ and $E_{s} = 4.8 \times 10^{-6} kg  s^{-2}$.  Experimental data points are the most frequent states observed at those parameters.Green denotes tank treading slipper, magenta denotes croissant shape, red denotes tumbling, blue denotes croissant and slipper bistability (croissant at low $Y_{0}$ and slipper at high $Y_{0}$) and grey denotes bistability of croissant (low $Y_{0}$) and other asymmetric shapes (like tumbling shapes at high $Y_{0}$).  No bistability depending on initial position of the capsule is observed for $\lambda=1$. The dashed lines serve as a guide to the eye.}
\end{figure*}

In this section, we discuss the dynamics of capsule in confined Poiseuille flow with $\lambda=1$. Here, the outward migration observed in unconfined flow is opposed by wall effects that push the capsule towards the center. The interplay between inward and outward migration can lead to different dynamics compared to the unconfined flow. We observe that for low confinement ratio with low capillary numbers, the wall push towards the center seems to cancel the outward migration and we observe tank treading slipper shapes. As the capillary number is increased (while keeping the confinement ratio fixed), a transition from slipper to croissant shape takes place as observed in unconfined flow. At high confinements, for example at $C_{n} = 0.65$, only croissant shapes are observed indicating the strong dominance of the wall push as confinement is increased. We observe no bistability in this case even though outward migration was observed in unconfined flow with $\lambda=1$. This indicates that the outward migration tendency is rather weak at $\lambda=1$ and gets cancelled by the wall push completely.  The results are presented in the form of a phase diagram in the parameter space of confinement ratio $C_{n}$ vs capillary number $C_{k}$ in Fig. \ref{fig:phasecapcon1}.  

 \subsection{\label{sec:conpoiseuilleflow} Confined Poiseuille flow $\lambda=5$ }
For confined Poiseuille flow with $\lambda=5$, we get a different picture. For low confinements, for example at $C_{n} = 0.35$, we observe tumbling dynamics at low capillary numbers ($C_{k} < 0.15$) and slippers at high capillary numbers ($ 0.15 \leq C_{k} \leq 0.5$). As the capillary number is increased ($C_{k} > 0.5$ ), we observe a croissant and slipper bistability, \emph{i.e.}, croissants are observed for low initial positions ($Y_{0}/R_{0} \leq 0.09$) and slippers at higher  initial positions ($Y_{0}/R_{0} > 0.09$).  This is inline with our conjecture that outward migration is counter balanced by the wall effects resulting in slipper shapes, especially when the initial position position of the capsule is further away from the centerline ($Y_{0}/R_{0}$ is large) . At high confinements $C_{n} > 0.5$, we observe the croissant-slipper bistability for high capillary numbers ($C_{k} \geq 0.23$) and croissant-tumbling bistability for low $C_{k}$.  The croissants are observed at low initial positions and slipper/tumbling at high initial positions. We combine these results in the form of a phase digram in the parameter space of confinement ratio $C_{n}$ vs capillary number $C_{k}$ as shown in Fig. \ref{fig:phasecapcon5}. 

Our shapes and results for confined flow are in good quantitative agreement with the experimental results. The shapes obtained are similar to the ones observed in \cite{41} (see Fig. \ref{fig:slipper} and \ref{fig:croissant}). The croissant-slipper bistability regime observed in experiments \cite{41, 30} is in agreement with our simulations (see Fig. \ref{fig:phasecapcon5} ) and could explain the existence of slippers at high capillary numbers. 

%
%

\section{Conclusion}
In  this  paper,  we study the dynamics and equilibrium shapes of 3D capsules in confined and unconfined Poiseuille flow with and without viscosity contrast.  We provide the phase  diagrams  for  both  the  cases. Our results for unconfined Poiseuille flow indicate the existence of outward migration tendency at both $\lambda=1$ and $\lambda=5$ unlike vesicles where the outward migration was only observed for $\lambda=5$. The croissant-slipper bistability in confined Poiseuille flow due to this outward migration is observed only for $\lambda=5$. The reason for this could be that the outward migration tendency is  weak for $\lambda=1$. As speculated in \cite{22}, the bistability for $\nu=0.65$ capsules carries over to confinement ratios $C_{n} > 0.5$ as well which was not the case with $\nu =0.9$ vesicles. Our results are in good agreement with experimental observations and extend the results for bistability observed in vesicles to capsules (a more realistic model for RBCs). Our results also provide good evidence for the validity of the speculation about bistability being the reason for experimental observation of slippers \cite{29, 41} at higher capillary numbers. We were able to reproduce results from several experimental and numerical prior works and we confirmed that the observed dynamics require a non-trivial, zero-stress, reference configuration space.

\end{document}